\documentclass[a4paper,12pt,titlepage]{article}

\usepackage{amsmath,amssymb,revsymb}
\usepackage{graphicx}
\usepackage{here}
\usepackage{bm}
\usepackage{fancybox}
\usepackage{secdot}
\usepackage{comment}
\usepackage{braket}
\usepackage{titlesec}
\usepackage{geometry}
\usepackage{hyperref}
\usepackage{slashed}
\usepackage{fancyhdr}
\usepackage{boxedminipage}
\usepackage{color}
\usepackage{feynmf}
\usepackage{cite}

\titleformat*{\section}{\LARGE\bfseries}
\titleformat*{\subsection}{\Large\bfseries}

\hypersetup{
setpagesize=false,
 bookmarksnumbered=true,%
 bookmarksopen=true,%
 colorlinks=true,%
 linkcolor=blue,
 citecolor=blue,
 }

\begin{document}


\fancypagestyle{foot}
{
\fancyfoot[L]{$^{*}$E-mail address : yakkuru\_111@ruri.waseda.jp \\ $^{\dagger}$E-mail address : tshd-sbsk.2015@ruri.waseda.jp}
\fancyfoot[C]{}
\fancyfoot[R]{}
\renewcommand{\headrulewidth}{0pt}
\renewcommand{\footrulewidth}{0.5pt}
}

\renewcommand{\footnoterule}{%
  \kern -3pt
  \hrule width \columnwidth
  \kern 2.6pt}


\begin{titlepage}
\begin{flushright}
\begin{minipage}{0.2\linewidth}
\normalsize


WU-HEP-19-09\\*[50pt]
\end{minipage}
\end{flushright}

\begin{center}

\vspace*{5truemm}
\Large
\bigskip\bigskip

\LARGE\textbf{Wilson-line Scalar as a Nambu-Goldstone Boson in Flux Compactifications and Higher-loop Corrections}

\Large

\bigskip\bigskip
Masaki Honda$^{1,*}$ and Toshihide Shibasaki$^{1,\dagger}$
\vspace{1cm}

{\large$^{1}$ \it{Department of Physics, Waseda University, Tokyo 169-8555, Japan}}\\
\bigskip\bigskip\bigskip

\large\textbf{Abstract}\\
\end{center}
We study a scalar zero mode originated from extradimensional components of a gauge field in a six-dimensional theory compactified on a magnetized torus. We confirm it is a Nambu-Goldstone boson of the translational symmetry on the torus which is breaking spontaneously due to magnetic flux. We also show explicitly it is massless up to the two-loop level. Moreover, we discuss full order contributions by considering the effective potential.

\thispagestyle{foot} 

\end{titlepage}

\baselineskip 7.5mm


\tableofcontents


\parindent=20pt

\section{Introduction}

The Standard Model (SM) is the most successful model in particle physics. However, there are still many mysteries in the SM. For example, the origin of the chiral structure and the generation structure of fermions has not been revealed. These suggest the existence of physics beyond the SM.

Extradimensional models are one of the most interesting possibilities beyond the SM. In particular, flux compactifications provide a lot of attractive results. In fact, four-dimensional (4D) chiral fermions and their generation structure can be realized by introducing magnetic fluxes in extradimensions \cite{Witten:1984dg}. Moreover, Yukawa couplings were computed explicitly in the magnetized models compactified on tori or toroidal orbifolds \cite{Cremades:2004wa, Abe:2008sx, Kobayashi:2010an, Matsumoto:2016okl}.

In extradimensional models, extradimensional components of gauge fields, which are often called Wilson-line (WL) scalars, are important since they can be candidates of 4D Higgs fields \cite{Fairlie:1979at, Manton:1979kb, Hosotani:1983xw, Hosotani:1988bm}. Recently, quantum corrections to the masses of WL scalars were calculated in flux compactifications. According to ref.  \cite{Buchmuller:2016gib, Ghilencea:2017jmh, Buchmuller:2018eog, Hirose:2019ywp}, the quantum corrections vanish at the one-loop level with and without supersymmetry. This result is confirmed in a six-dimensional (6D) U(1) gauge theory \cite{Buchmuller:2018eog} and an SU(2) gauge theory \cite{Hirose:2019ywp} compactified on a torus $T^2$ with magnetic flux. In the previous researches, the physical reason for the cancellation of the mass corrections has been under discussion: the shift symmetry of the WL scalars \cite{Buchmuller:2018eog} or Nambu-Goldstone (NG) bosons of the translational symmetry on a torus \cite{Buchmuller:2016gib, Hirose:2019ywp}.

In this paper, we confirm in more detail that the WL scalar is an NG boson of the translational symmetry by considering commutation relations between the WL scalar and the momentum operators. We focus on a 6D U(1) gauge theory with a single Weyl fermion compactified on a magnetized torus as the simplest case. We calculate quantum corrections to the mass of the WL scalar and show the cancellation up to the two-loop level. To confirm its masslessness in the full order, we mention the effective potential does not depend on the value of the WL phase in our setup.

This paper is organized as follows. In section \ref{sec.2}, we explain the U(1) gauge theory on a magnetized torus. In section \ref{sec.3}, we show the WL scalar is an NG Boson of the translations. In section \ref{sec.4}, we calculate quantum corrections to the mass of the WL scalar. We summarize in section \ref{sec.5}.

\section{Gauge Theory on a Magnetized Torus}
\label{sec.2}

We consider a 6D U(1) gauge theory on $M^4 \times T^2$ with magnetic flux. In the following, we consider a square torus as extradimensions and the coordinates on the torus take values in the interval $x_5,x_6 \in [0,1)$ for simplicity. We denote 6D spacetime indices by $M = 0, \cdots ,3,5,6$, 4D indices by $\mu = 0, \cdots ,3$ and extradimensional indices by $m = 5,6$. We take the metric as $g_{MN} = \mathrm{diag}(-1,+1, \cdots , +1)$, and define complex coodinates
\begin{align}
	z = \frac{1}{2}(x_5+ix_6) , \ \ \ \ \ \ \ \ 
	\partial_z = \partial_5 - i \partial_6 .
\end{align}

We introduce a left-handed 6D Weyl fermion $\Psi$ interacting with a U(1) gauge field $A_M$. The Lagrangian is given by
\begin{align}
\label{6D-Lag.1}
	\mathcal{L} = -\frac{1}{4}F^{MN}F_{MN} + i\overline{\Psi}\Gamma^MD_M\Psi - \frac{1}{2\xi}\left(\partial_\mu A^\mu + \xi\partial_m A^m\right)^2 ,
\end{align}
where  $F_{MN} = \partial_M A_N - \partial_N A_M $, $D_M = \partial_M + iqA_M $ and $\Gamma^7\Psi = -\Psi $. The third term is a gauge fixing term and $\xi \geq 0$ is a gauge fixing parameter.\footnote{This is an extension of the $R_\xi$ gauge in the SM.} We summarize 6D gamma matrices in appendix \ref{app.A}. It is convenient to define a complex field $\phi$ as
\begin{align}
	\phi = \frac{1}{\sqrt{2}}\left(A_6 + iA_5\right) .
\end{align}

A constant magnetic flux $f$ is given by the nontrivial background or VEVs of extradimensional components of the gauge field. In this paper, we introduce it as follows.
\begin{align}
\label{flux1}
	\langle A_5 \rangle = -\frac{1}{2} f \left(x_6 + \theta_6 \right) ,\ \ \ \ \ \ \ \ 
	\langle A_6 \rangle = \frac{1}{2} f \left(x_5 + \theta_5 \right) ,
\end{align}
where $\theta_{5,6}$ are constant WL phases. This configuration satisfies the classical equation of motion
\begin{align}
	\partial^M \langle F_{MN} \rangle = 0 .
\end{align}
Accordingly, the VEV of the complex field $\phi$ is given as
\begin{align}
	\langle\phi\rangle = \frac{1}{\sqrt{2}} f \left(\overline{z} + \overline{\theta}\right) ,
\end{align}
where $\theta = \left(\theta_5 + i\theta_6 \right)/2$. The magnetic flux is quantized on the torus,
\begin{align}
	\frac{q}{2\pi} \int_{T^2} d^2x \langle F_{56} \rangle = \frac{qf}{2\pi} = N \in \mathbb{Z} .
\end{align}

Using two-component Weyl fermions $\psi$ and $\chi$ defined in appendix \ref{app.A}, we can rewrite the Lagrangian (\ref{6D-Lag.1}) as follows.
\begin{align}
\label{6D-Lag.2}
	\mathcal{L} =& -\frac{1}{4}F^{\mu\nu}F_{\mu\nu} -\frac{1}{2\xi}\partial_\mu A^\mu \partial_\nu A^\nu - \partial_\mu \varphi^\dagger \partial^\mu\varphi - \frac{1}{2} \partial_{\overline{z}}A^\mu\partial_zA_\mu -\frac{1}{2}f^2 \nonumber \\
		&  - \frac{1}{4}\left(\partial_z \varphi^\dagger + \partial_{\overline{z}}\varphi\right)^2 +\frac{\xi}{4}\left(\partial_z \varphi^\dagger - \partial_{\overline{z}}\varphi \right)^2 + i\psi\sigma^\mu\overline{D}_\mu \psi^\dagger + i\chi \sigma^\mu D_\mu \chi^\dagger \nonumber \\
		& + \chi\left(\partial_z + 2\pi N\left(\overline{z} + \overline{\theta}\right) + \sqrt{2}q\varphi\right)\psi + \chi^\dagger\left(\partial_{\overline{z}} + 2\pi N\left(z + \theta\right) + \sqrt{2}q\varphi^\dagger\right)\psi^\dagger ,
\end{align}
where $\varphi$ is a quantum fluctuation of $\phi$ around the VEV,
\begin{align}
	\phi = \langle \phi \rangle + \varphi .
\end{align}

In the following of this paper, we consider $N > 0$ without loss of generality. We define following operators,
\begin{align}
	a &= \frac{1}{\sqrt{4\pi N}}\left(\partial_z + 2\pi N\left(\overline{z} + \overline{\theta}\right)\right) , \\
	a^\dagger &= - \frac{1}{\sqrt{4\pi N}}\left(\partial_{\overline{z}} - 2\pi N\left(z + \theta\right)\right) .
\end{align}
These satisfy the commutaion relations of creation and annihilation operators,
\begin{align}
	[a,a^\dagger] =1 ,\ \ \ \ \ \ \ \ 
	\mathrm{others} = 0 .
\end{align}
Mode functions of the two-component fermions are defined by these operators as
\begin{align}
	a\lambda_{0j}(z + \theta) = 0 ,\ \ \ \ \ \ \ \ 
	\lambda_{nj}(z + \theta) = \frac{1}{\sqrt{n!}}\left(a^\dagger\right)^n\lambda_{0j}(z + \theta) .
\end{align}
where $j = 0 , \cdots , |N|-1$ labels the degeneracy.\footnote{Explicit forms of the mode functions including massive modes are studied in ref.\cite{Hamada:2012wj}.} The WL phase $\theta$ shifts the position of the mode functions on the torus. These functions satisfy a orthonormality condition
\begin{align}
	\int_{T^2}d^2x \overline{\lambda}_{nj}(z+\theta)\lambda_{n'j'}(z+\theta) = \delta_{nn'}\delta_{jj'} .
\end{align}
By use of this set of functions, the 6D fermions are Kaluza-Klein (KK) expanded,
\begin{align}
\label{KK-psi}
	\psi(x,z) &= \sum_{n,j}\psi_{nj}(x)\lambda_{nj}(z+\theta) , \\
\label{KK-xi}
	\chi(x,z) &= \sum_{n,j}\chi_{nj}(x)\overline{\lambda}_{nj}(z+\theta) .
\end{align}
It is useful to define 4D left-handed Weyl fermions $\Psi_{0j}$ and Dirac fermions $\Psi_{nj}$,
\begin{align}
	\Psi_{0j} = \left(\begin{array}{c} \psi_{0j} \\ 0 \end{array} \right) ,\ \ \ \ \ \ \ \ 
	\Psi_{nj} = \left(\begin{array}{c} \psi_{nj} \\ \chi^\dagger_{n-1j} \end{array} \right) .
\end{align}

The KK expansions for bosons are given by
\begin{align}
\label{KK-A}
	A_\mu(x,z) &= \sum_{l,m} A_{\mu,lm}(x)e^{zM_{lm}-\overline{z}\overline{M}_{lm}} , \\
\label{KK-phi}
	\varphi(x,z) &= \sum_{l,m} \varphi_{lm}(x)e^{zM_{lm}-\overline{z}\overline{M}_{lm}} ,
\end{align}
where $M_{lm} = 2\pi\left(m+il\right)$. $A_{\mu,-l-m} = A^\dagger_{\mu,lm}$ is satisfied since the gauge field $A_\mu$ is real. The scalar zero mode $\varphi_0$ is called a WL scalar because the WL phase multiplied by an appropriate factor $f \overline{\theta}/\sqrt{2}$ can be interpreted as its VEV. Before the dimensional reduction, we define 4D complex scalar fields $\eta_{lm}$ and $\sigma_{lm}$ when $l$ or $m$ is non-zero for convenience,
\begin{align}
	\eta_{lm} &= \frac{|M_{lm}|}{\sqrt{2}}\left(\frac{1}{M_{lm}}\varphi_{lm} - \frac{1}{\overline{M}_{lm}}\varphi^\dagger_{-l-m}\right) , \\
	\sigma_{lm} &= \frac{|M_{lm}|}{\sqrt{2}}\left(\frac{1}{M_{lm}}\varphi_{lm} + \frac{1}{\overline{M}_{lm}}\varphi^\dagger_{-l-m}\right) .
\end{align}

Substituting the KK expansions (\ref{KK-psi}), (\ref{KK-xi}), (\ref{KK-A}) and (\ref{KK-phi}) into the Lagrangian (\ref{6D-Lag.2}) and integrating it on a torus, we obtain the following 4D effective Lagrangian,
\begin{align}
\label{4D-Lag.}
	\mathcal{L}_{4D} =& - \frac{1}{4}F^{\mu\nu}_0 F_{\mu\nu,0} - \frac{1}{2\xi}\partial_\mu A^\mu_0 \partial_\nu A^\nu_0 \nonumber \\
		& - \sum_{l,m} \left(\frac{1}{2}F^{\mu\nu\dagger}_{lm} F_{\mu\nu,lm} + \frac{1}{\xi}\partial_\mu A^{\mu\dagger}_{lm} \partial_\nu A^\nu_{lm} + |M_{lm}|^2A^{\mu\dagger}_{lm}A_{\mu,lm} \right) - \frac{1}{2}f^2 \nonumber \\
		&  - \partial_\mu \varphi^\dagger_0 \partial^\mu \varphi_0 - \sum_{l,m} \left(\partial_\mu \eta^\dagger_{lm} \partial^\mu \eta_{lm} + \partial_\mu \sigma^\dagger_{lm} \partial^\mu \sigma_{lm} + |M_{lm}|^2\eta^\dagger_{lm}\eta_{lm} + \xi |M_{lm}|^2\sigma^\dagger_{lm}\sigma_{lm} \right) \nonumber \\
		& + \sum_{n,j}\left(i\overline{\Psi}_{nj}\gamma^\mu \partial_\mu \Psi_{nj} + \sqrt{4\pi Nn}\overline{\Psi}_{nj}\Psi_{nj} \right) - \sum_{n,j} q\overline{\Psi}_{nj} \gamma^\mu A_{\mu,0} \Psi_{nj} \nonumber \\
		& - \sum_{l,m;n,j;n',j'} \left(qe^{-\theta M_{lm} + \overline{\theta} \overline{M}_{lm}} \overline{\Psi}_{nj} \gamma^{\mu,lm}_{nj,n'j'} A_{\mu,lm} \Psi_{n'j'} + qe^{\theta M_{lm} - \overline{\theta} \overline{M}_{lm}} \overline{\Psi}_{nj} \gamma^{\mu,-l-m}_{nj,n'j'} A^\dagger_{\mu,lm} \Psi_{n'j'} \right) \nonumber \\
		&  + \sum_{n,j} \left(\sqrt{2}q\varphi_0 \overline{\Psi}_{n+1j} P_L \Psi_{nj} +\sqrt{2}q\varphi^\dagger_0 \overline{\Psi}_{nj} P_R \Psi_{n+1j} \right) \nonumber \\
		& + \sum_{l,m;n,j;n',j'} \left\{\frac{q}{|M_{lm}|} e^{-\theta M_{lm} + \overline{\theta} \overline{M}_{lm}} \left(\eta_{lm} \overline{\Psi}_{nj} A^{lm}_{nj,n'j'} \Psi_{n'j'} + \sigma_{lm} \overline{\Psi}_{nj} B^{lm}_{nj,n'j'} \Psi_{n'j'} \right) \right. \nonumber \\
		& \left. + \frac{q}{|M_{lm}|} e^{\theta M_{lm} - \overline{\theta} \overline{M}_{lm}} \left(\eta^\dagger_{lm} \overline{\Psi}_{nj} A^{-l-m}_{nj,n'j'} \Psi_{n'j'} - \sigma^\dagger_{lm} \overline{\Psi}_{nj} B^{-l-m}_{nj,n'j'} \Psi_{n'j'} \right) \right\} ,
\end{align}
where $P_{L,R}$ are 4D chirality projection operators
\begin{align}
	P_L = \frac{1-\gamma_5}{2} ,\ \ \ \ \ \ \ \ 
	P_R = \frac{1+\gamma_5}{2} ,
\end{align}
and the coefficients are given by
\begin{align}
	\gamma^{\mu,lm}_{nj,n'j'} &= \left(\begin{array}{cc} 0 & C^{lm}_{n-1j,n'-1j'} \sigma^\mu \\ C^{lm}_{nj,n'j'} \overline{\sigma}^\mu & 0 \end{array} \right) , \\
	A^{lm}_{nj,n'j'} &= \left(\begin{array}{cc} M_{lm} C^{lm}_{n-1j,n'j'} & 0 \\ 0 & -\overline{M}_{lm} C^{lm}_{nj,n'-1j'} \end{array} \right) , \\
	B^{lm}_{nj,n'j'} &= \left(\begin{array}{cc} M_{lm} C^{lm}_{n-1j,n'j'} & 0 \\ 0 & \overline{M}_{lm} C^{lm}_{nj,n'-1j'} \end{array} \right) ,
\end{align}
with the overlap integrals
\begin{align}
	C^{lm}_{nj,n'j'} = \int_{T^2}d^2x e^{zM_{lm} - \overline{z}\overline{M}_{lm}} \overline{\lambda}_{nj}(z) \lambda_{n'j'}(z) .
\end{align}
When $n$ or $n'$ is a negative integer, we define $C^{lm}_{nj,n'j'}=0$. Here, $C^{lm}_{nj,n'j'}$ is defined for the WL phase $\theta=0$ and independent of $\theta$. In the 4D effective Lagrangian (\ref{4D-Lag.}) and the following, the sum of $l$ and $m$ is in the following range,
\begin{align}
	-\infty \leq l \leq \infty ,\ \ \ \ \ \ \ \ 
	\begin{cases}
		0 \leq m \leq \infty & (l>0) \\
		1 \leq m \leq \infty & (l \leq 0) .
	\end{cases}
\end{align}
We note the scalar fields $\eta_{lm}$ and $\sigma_{lm}$ are physical particles and NG bosons in the 4D effective theory, respectively. The KK masses of NG bosons $\sigma_{lm}$ depend on the gauge fixing parameter $\xi$ and they are eaten by the massive modes of the 4D vector fields $A_{\mu,lm}$ as their longitudinal components.

By the definition of the overlap integrals, $\overline{C}^{lm}_{nj,n'j'} = C^{-l-m}_{n'j',nj}$ holds. The following recurrence relations are obtaind from the definition of the mode functions of the fermions,
\begin{align}
\label{recurrence-1}
	\sqrt{4\pi N} \left(\sqrt{n} C^{lm}_{n-1j,n'j'} - \sqrt{n'+1} C^{lm}_{nj,,n'+1j'} \right) &= \overline{M}_{lm} C^{lm}_{nj,n'j'} , \\
\label{recurrence-2}
	\sqrt{4\pi N} \left(\sqrt{n+1} C^{lm}_{n+1j,n'j'} - \sqrt{n'} C^{lm}_{nj,,n'-1j'} \right) &= M_{lm} C^{lm}_{nj,n'j'} .
\end{align}
These equations are useful to calculate two-loop corrections in section \ref{sec.4}.

\section{Wilson-line Scalar as an NG Boson}
\label{sec.3}

The Lagrangian (\ref{6D-Lag.2}) does not have the translational symmetry on the torus unlike the Lagrangian (\ref{6D-Lag.1}) since the magnetic flux is introduced as the VEVs of extradimensional components of the gauge field and they depends on the coordinates on the torus. However, the Lagrangian (\ref{6D-Lag.2}) is invariant under the transformation combining the translations on the torus and a constant shift for the fluctuation field $\varphi$ because of the original translational symmetry.
\begin{align}
	\delta A_\mu , \psi , \chi = \left(\epsilon \partial_z + \overline{\epsilon} \partial_{\overline{z}} \right) A_\mu , \psi , \chi ,\ \ \ \ \ \ \ \ 
	\delta \varphi =  \left(\epsilon \partial_z + \overline{\epsilon} \partial_{\overline{z}} \right) \varphi + \frac{\overline{\epsilon}}{\sqrt{2}} f .
\end{align}
Especially, this shift symmetry of the 6D field $\varphi$ is rewritten to that of the WL scalar $\varphi_0$ in the 4D effective theory \cite{Buchmuller:2018eog}. Therefore, the WL scalar in this setup is expected to be an NG boson of the translational symmetry.

To confirm this in more detail, let us consider commutation relations between the fluctuation $\varphi$ and the Noether charges $P_{z, {\overline{z}}}$ of the translations on the torus. Deriving the energy momentum tensor from the Lagrangian before introducing the magnetic flux, we obtain following terms containing $\varphi$,
\begin{align}
\label{stress1}
	{T^0}_z & \ni \frac{\partial \mathcal{L}}{\partial(\partial_0 \phi)} \partial_z \phi + \frac{\partial \mathcal{L}}{\partial(\partial_0 \phi^\dagger)} \partial_z \phi^\dagger
		= - \partial^0 \varphi^\dagger \partial_z \varphi - \partial^0 \varphi \partial_z \varphi^\dagger - \frac{f}{\sqrt{2}} \partial^0 \varphi , \\
\label{stress2}
	{T^0}_{\overline{z}} & \ni \frac{\partial \mathcal{L}}{\partial(\partial_0 \phi)} \partial_{\overline{z}} \phi + \frac{\partial \mathcal{L}}{\partial(\partial_0 \phi^\dagger)} \partial_{\overline{z}} \phi^\dagger
		= - \partial^0 \varphi^\dagger \partial_{\overline{z}} \varphi - \partial^0 \varphi \partial_{\overline{z}} \varphi^\dagger - \frac{f}{\sqrt{2}} \partial^0 \varphi^\dagger .
\end{align}
Since $-\partial^0 \varphi^\dagger \ (-\partial^0 \varphi)$ is the conjugate momentum of $\varphi \ (\varphi^\dagger)$, the commutation relations between $P_{z,\overline{z}}$ and $\varphi$ are given as follows.
\begin{align}
	[iP_{\overline{z}},\varphi] & \equiv i \int d^3x \int_{T^2}d^2x [{T^0}_{\overline{z}}(x),\varphi(y)] = \partial_{\overline{z}}\varphi + \frac{1}{\sqrt{2}}f , \\
	[iP_z,\varphi^\dagger] & \equiv i \int d^3x \int_{T^2}d^2x [{T^0}_z(x),\varphi^\dagger(y)] = \partial_z \varphi^\dagger + \frac{1}{\sqrt{2}}f .
\end{align}
The VEVs of them are
\begin{align}
\label{6D-commutator-VEV}
	\langle [iP_{\overline{z}},\varphi] \rangle = \frac{f}{\sqrt{2}} , \ \ \ \ \ \ \ \ 
	\langle [iP_z,\varphi^\dagger] \rangle = \frac{f}{\sqrt{2}} , \ \ \ \ \ \ \ \ 
	\mathrm{others} = 0 .
\end{align}
This result implies the fluctuation $\varphi$ is an NG boson of the translations.\footnote{The translational symmetry is assumed in the proof of the ordinary NG theorem. Thus, eqs. (\ref{6D-commutator-VEV}) may not be sufficient to claim the 6D field $\varphi$ is an NG boson of the translations. However, this is not a problem in the discussion about the WL scalar $\varphi_0$ because the translational symmetry in the 4D effective theory is not breaking.}

Then, we show the WL scalar $\varphi_0$ takes over the property of the 6D field $\varphi$ as an NG boson after the dimensional reduction. Substituting the KK expansion of $\varphi$ into eqs. (\ref{stress1}) and (\ref{stress2}) and integrating them on the torus, we obtain the momentum density in the 4D effective theory,
\begin{align}
	{{T_{4D}}^0}_z & = \int_{T^2}d^2x {T^0}_z \nonumber \\
		& \ni - \frac{f}{\sqrt{2}} \partial^0 \varphi_0  - \sum_{l,m} M_{lm} \left(\eta_{lm}\partial^0\eta^\dagger_{lm} -\eta^\dagger_{lm} \partial^0\eta_{lm} + \sigma_{lm}\partial^0\sigma^\dagger_{lm} -\sigma^\dagger_{lm} \partial^0\sigma_{lm} \right) , \\
	{{T_{4D}}^0}_{\overline{z}} & = \int_{T^2}d^2x {T^0}_{\overline{z}} \nonumber \\
		& \ni - \frac{f}{\sqrt{2}} \partial^0 \varphi^\dagger_0  - \sum_{l,m} \overline{M}_{lm} \left(\eta_{lm}\partial^0\eta^\dagger_{lm} -\eta^\dagger_{lm} \partial^0\eta_{lm} + \sigma_{lm}\partial^0\sigma^\dagger_{lm} -\sigma^\dagger_{lm} \partial^0\sigma_{lm} \right) .
\end{align}
Commutation relations between  $P_{z,\overline{z}}$ and $\varphi_0$ are
\begin{align}
	[iP_{\overline{z}},\varphi_0] & \equiv i \int d^3x [{{T_{4D}}^0}_{\overline{z}}(x),\varphi_0(y)] = \frac{f}{\sqrt{2}} , \\
	[iP_z,\varphi_0^\dagger] & \equiv i \int d^3x [{{T_{4D}}^0}_z(x),\varphi_0^\dagger(y)] = \frac{f}{\sqrt{2}} .
\end{align}
Therefore,
\begin{align}
	\langle [iP_{\overline{z}},\varphi_0] \rangle = \frac{f}{\sqrt{2}} , \ \ \ \ \ \ \ \ 
	\langle [iP_z,\varphi^\dagger_0] \rangle = \frac{f}{\sqrt{2}} , \ \ \ \ \ \ \ \ 
	\mathrm{others} = 0 .
\end{align}
In the 4D effectve theory, the WL scalar $\varphi_0$ is an NG boson of the translational symmetry on the torus.

\section{Quantum Corrections}
\label{sec.4}

In section \ref{sec.3}, we showed the WL scalar $\varphi_0$ is an NG boson. Therefore, it is expected to be massless in full order of the perturbation. In this section, we show it is true up to the two-loop level by diagrammatic calculations and in the full order from the WL phase independence of the effective potential. Feynman rules we use in the following are obtained from the 4D effective Lagrangian (\ref{4D-Lag.}).

\subsection{One-loop Corrections}

\begin{figure}[H]
\begin{center}
\includegraphics[width=5cm]{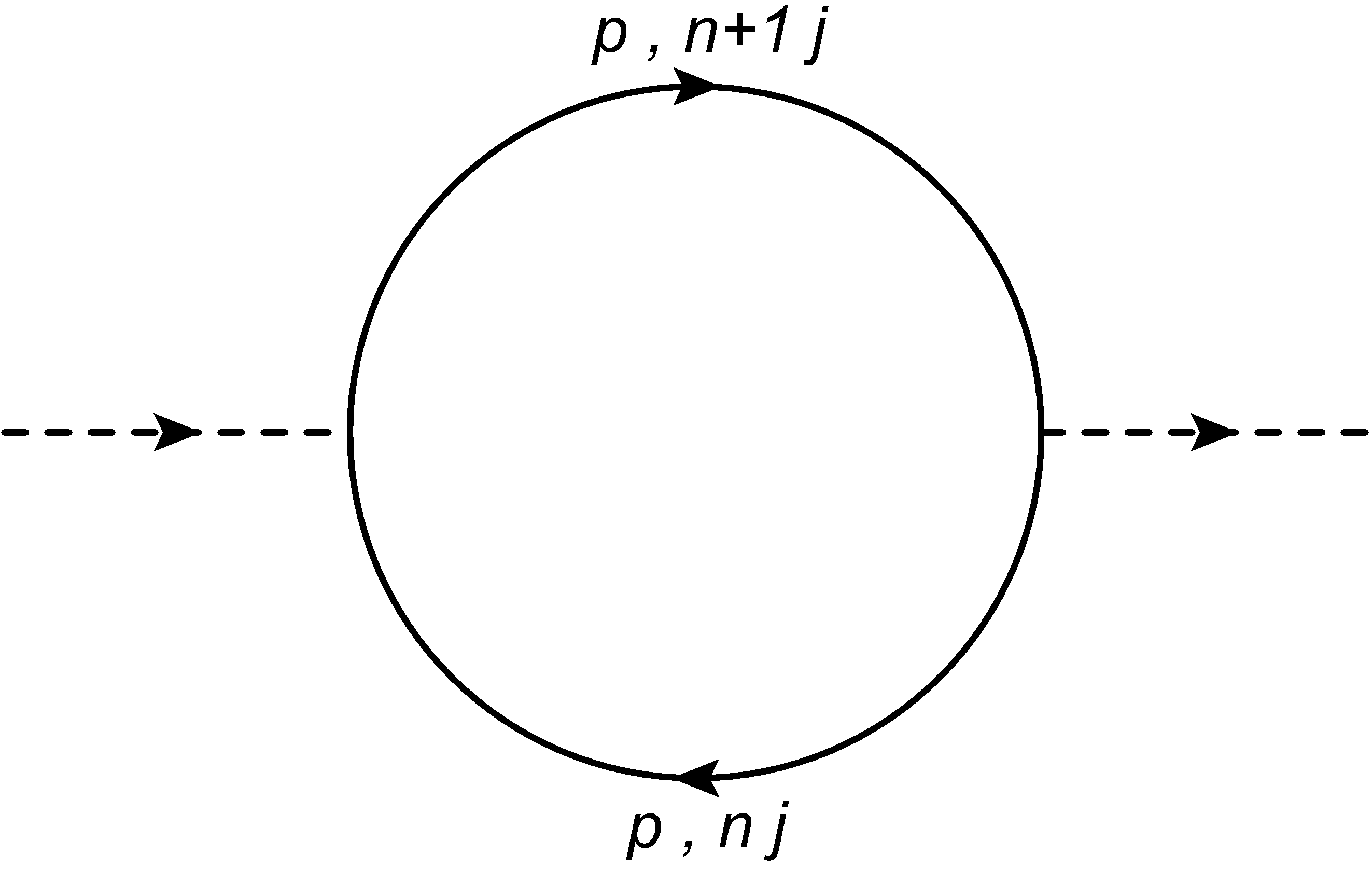}
\caption{One-loop contributions to the WL scalar mass. Indices of internal lines represent momentum and KK mode, respectively.}
\label{fig.1}
\end{center}
\end{figure}

First, we review the one-loop calculation \cite{Buchmuller:2016gib, Buchmuller:2018eog}. Fig. \ref{fig.1} shows one-loop contributions to the mass of $\varphi_0$. The sum of the loops over the mode of fermions is
\begin{align}
\label{one-loop}
	I^{\mathrm{1 \text{-} loop}} &= 16\pi Nq^2 \sum_{n=0}^\infty \int \frac{d^4p}{(2\pi)^4} \frac{p^2}{(p^2+n)(p^2+n+1)} \nonumber \\
		&= 16\pi Nq^2 \sum_{n=0}^\infty \int \frac{d^4p}{(2\pi)^4} \left(\frac{n+1}{p^2+n+1} - \frac{n}{p^2+n} \right) \nonumber \\
		&= 0 ,
\end{align}
where an appropriate change of the integral variable is performed to simplify the masses in propagators. In the second to third line, we regularize the sum by the shift $n\rightarrow n+1$ in the second term.\footnote{The same result is obtained by the combination of dimensional regularization and zeta function regularization \cite{Ghilencea:2017jmh}.} Therefore, the WL scalar $\varphi_0$ is massless at one-loop level.

If the sum is truncated in a finite number of terms, the correction is quadratically divergent.\footnote{Such a truncation for the sum over the mode of fermions does not preserve the translational symmetry on the torus because the eigenstates of $\partial_z$ or $\partial_{\overline{z}}$ are not the basis of the KK expansion \cite{Buchmuller:2018eog}.} Usually, such a divergence must be removed by counterterms. However, when we consider the entire KK tower and calculate like eqs. (\ref{one-loop}), no divergence appears. We also use this regularization method in the following two-loop calculations.

\subsection{Two-loop Corrections}

\begin{figure}[H]
\begin{center}
\includegraphics[width=5cm]{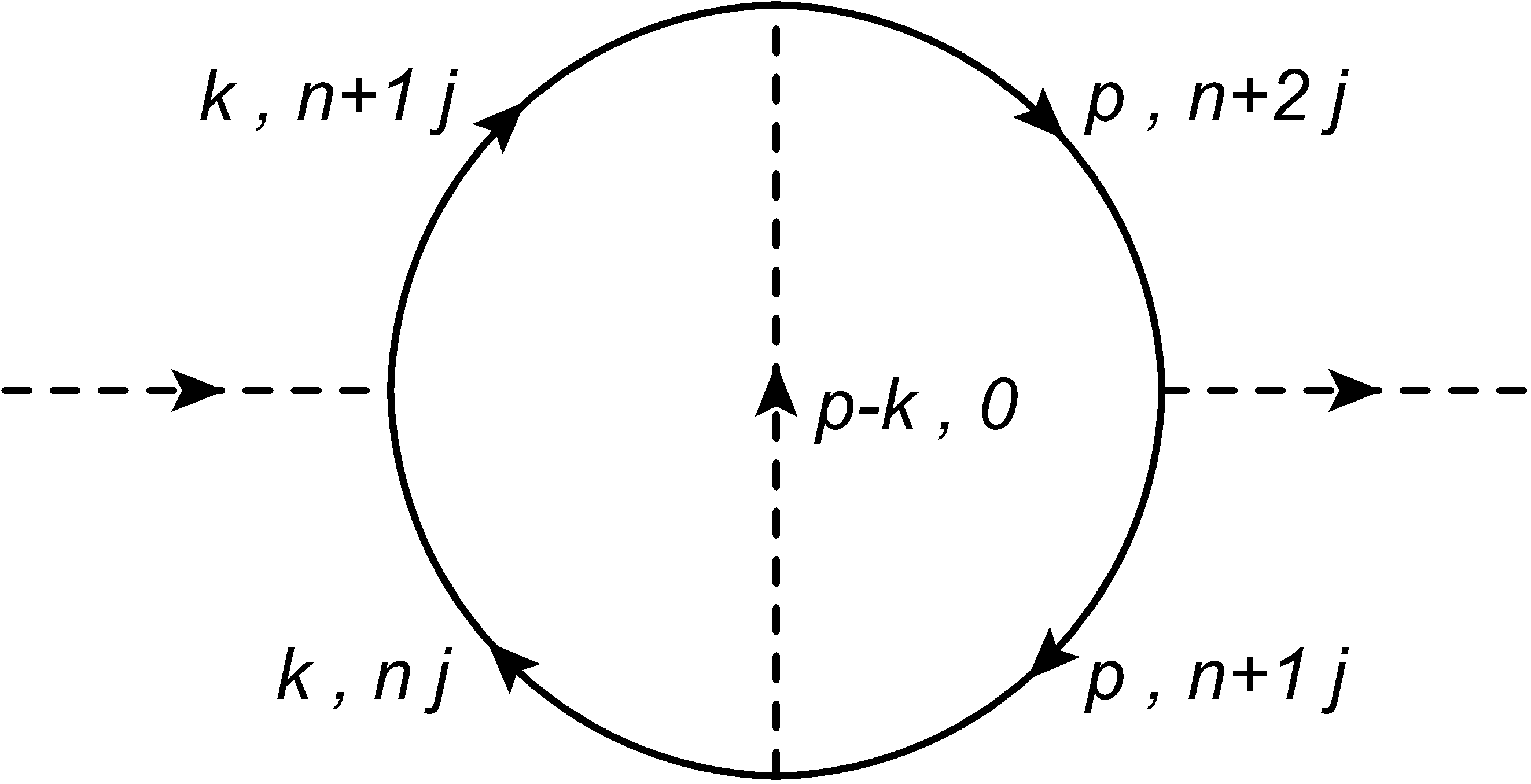} \ \ \ 
\includegraphics[width=5cm]{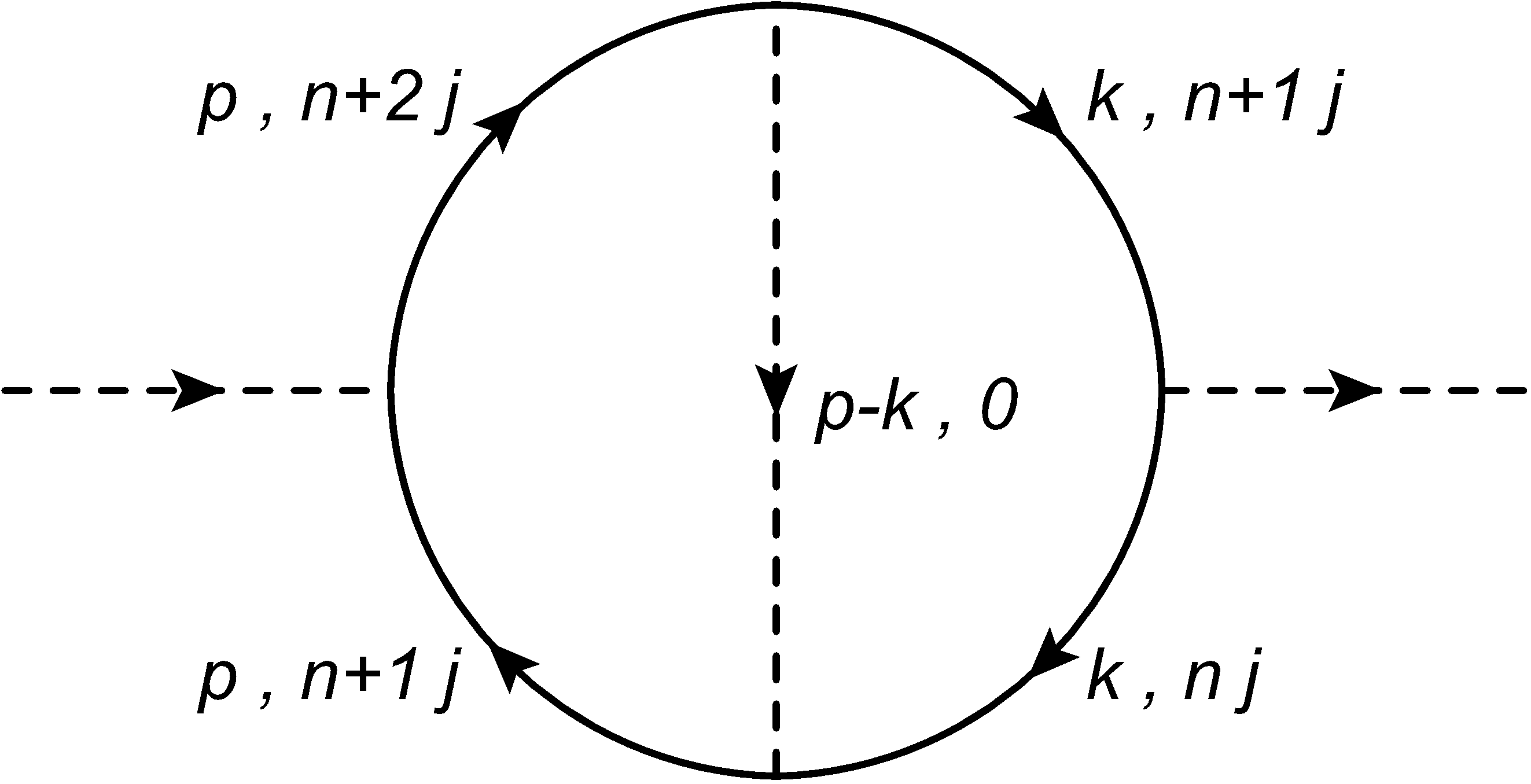} \ \ \ 
\mbox{\raisebox{-3mm}{\includegraphics[width=5cm]{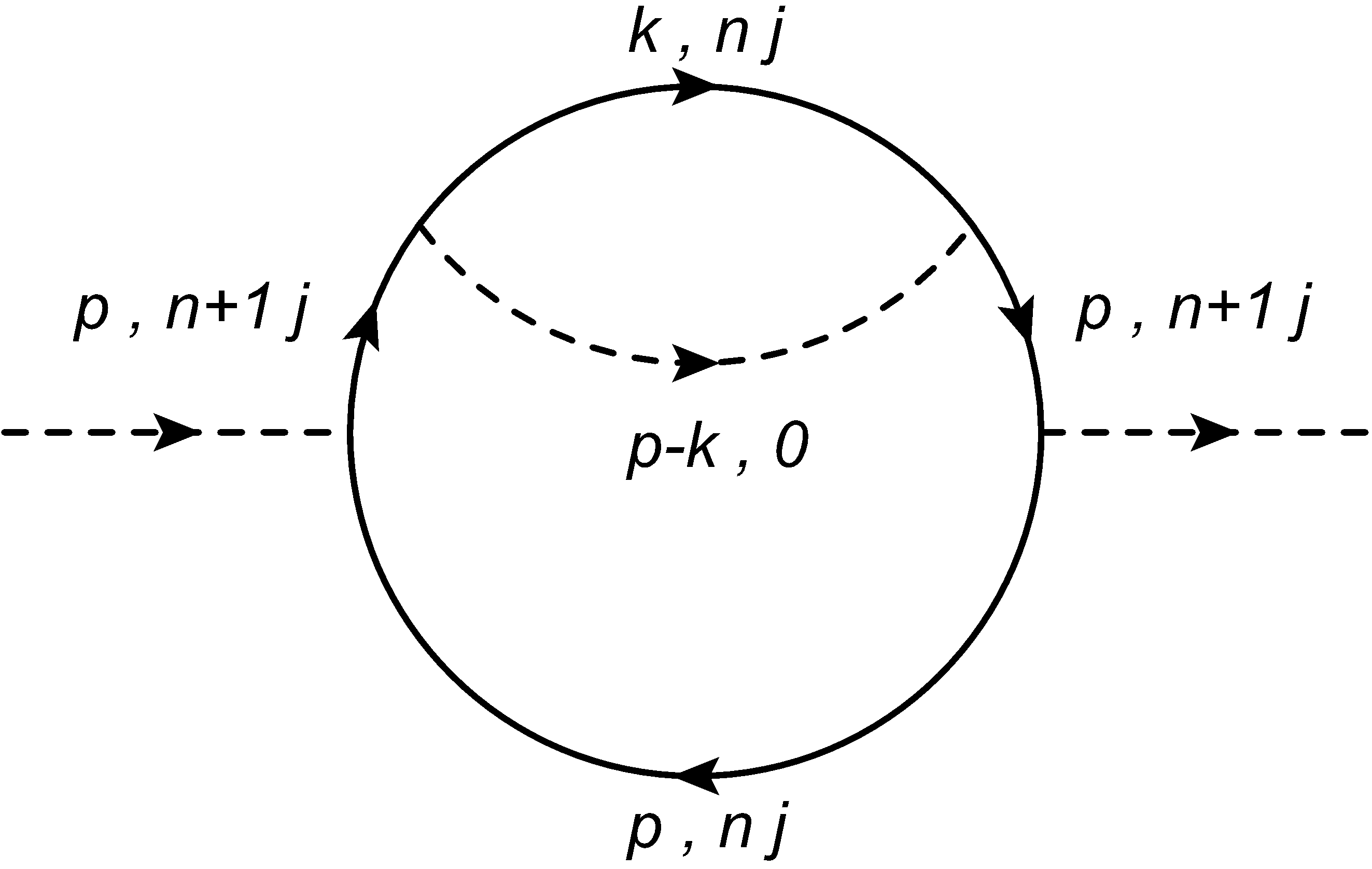}}} \\
\vspace{\baselineskip}
\includegraphics[width=5cm]{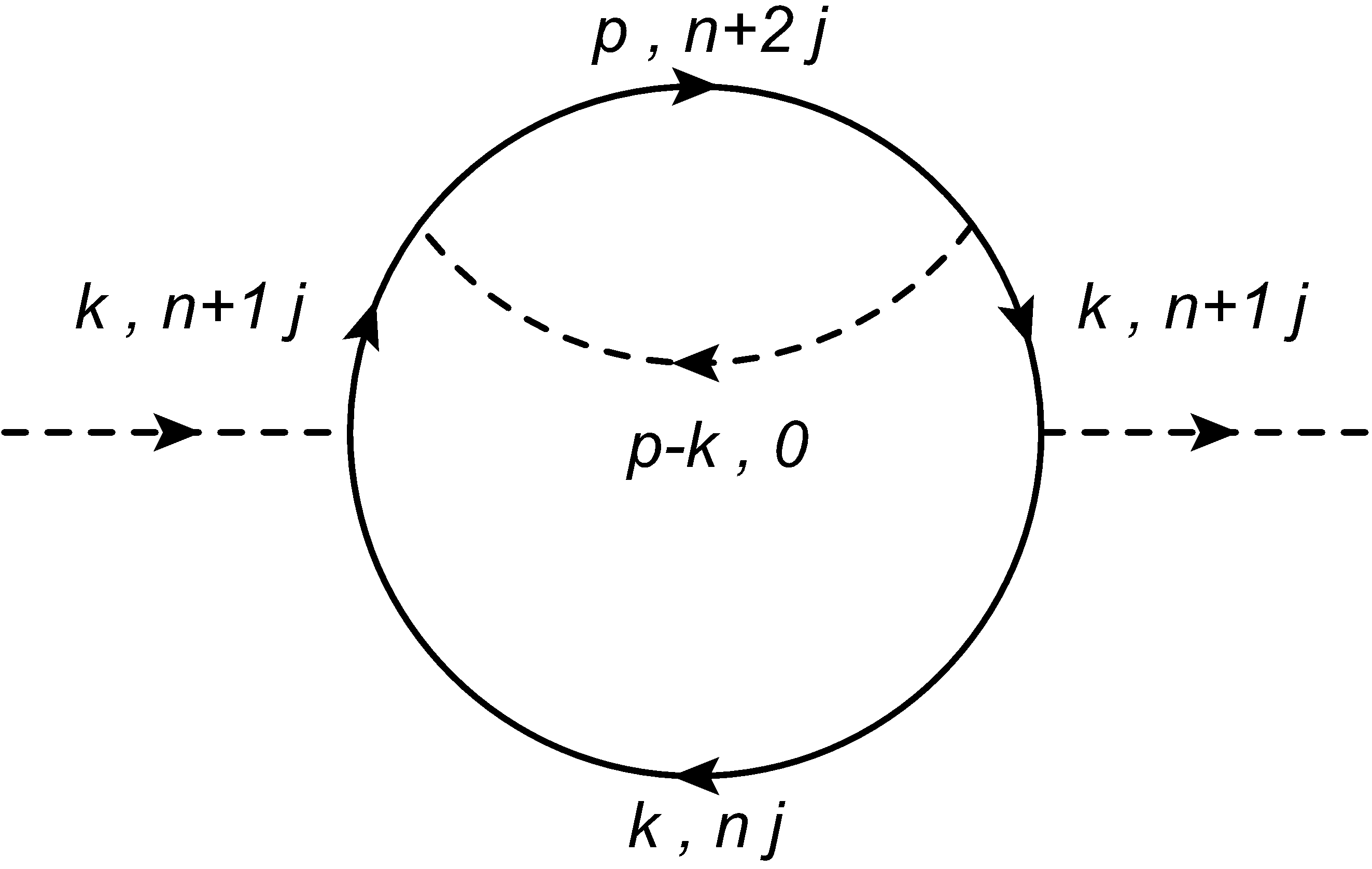} \ \ \ 
\includegraphics[width=5cm]{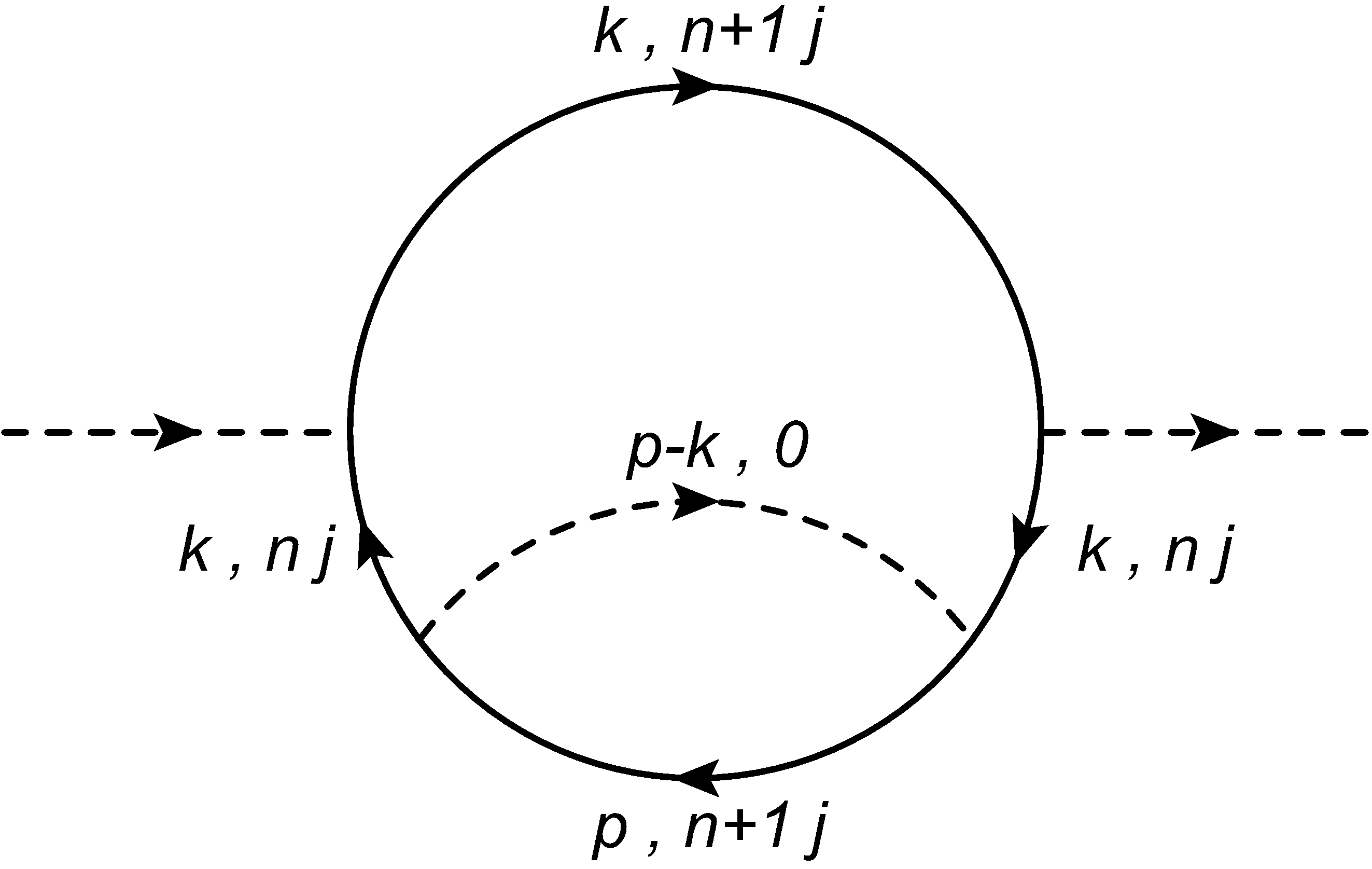} \ \ \ 
\includegraphics[width=5cm]{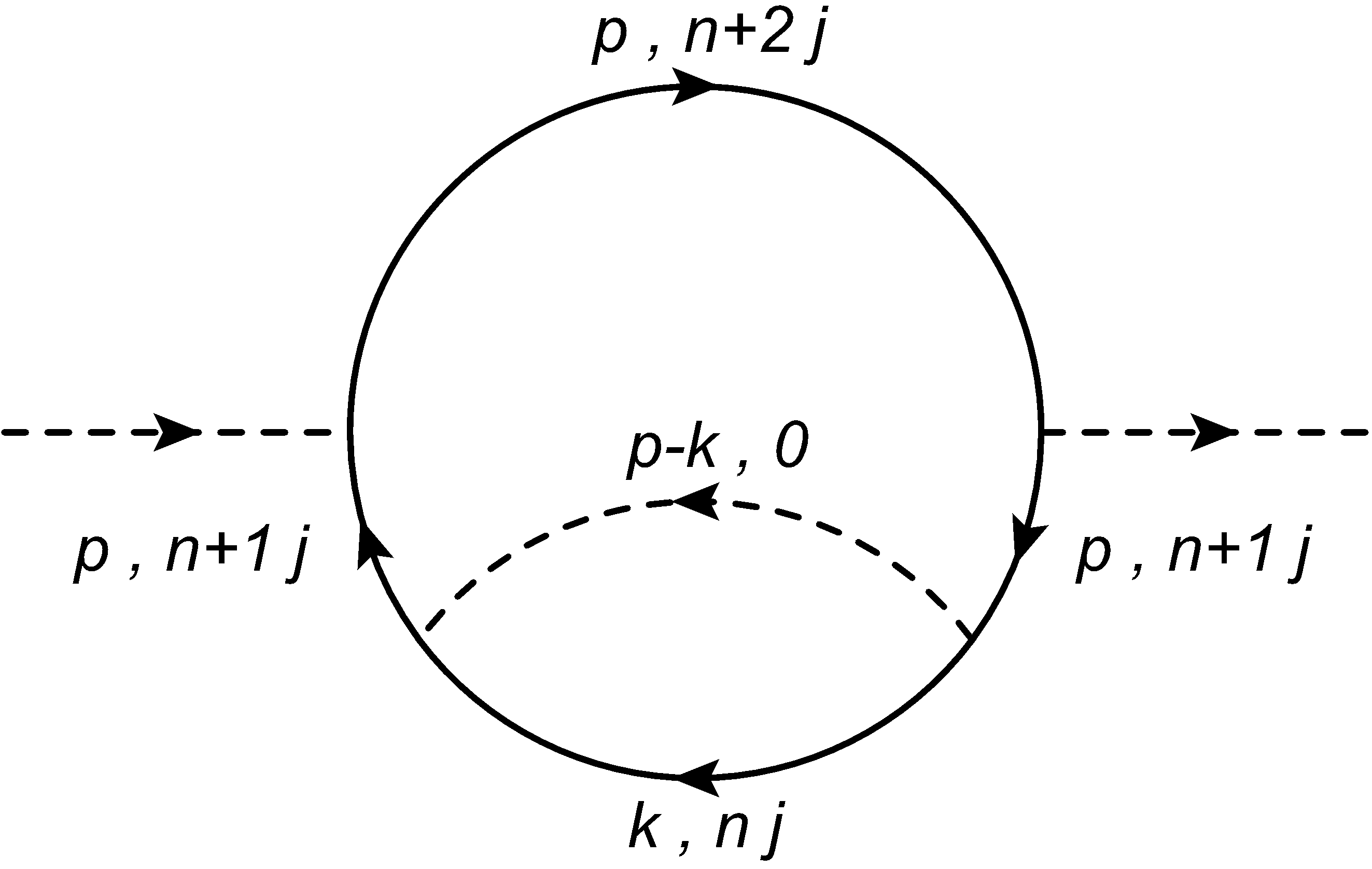}
\caption{Two-loop contributions including $\varphi_0$ as an internal line.}
\label{fig.2}
\end{center}
\end{figure}

Let us consider two-loop contributions to the mass of $\varphi_0$. There are six types of two-loop contributions including the WL scalar $\varphi_0$ as an internal line (see fig. \ref{fig.2}). The sum of these two-loops over the mode of fermions is
\begin{align}
	I^{\mathrm{2 \text{-} loop}}_{\varphi_0} = & -32 \pi i N q^4 \sum_{n=0}^{\infty} \int \frac{d^4p d^4k}{(2 \pi)^8} \frac{p \cdot k}{(p-k)^2} \nonumber \\
			& \times \left[\frac{2(n+1)}{(p^2+n+1)(p^2+n+2)(k^2+n)(k^2+n+1)} \right. \nonumber \\
			& \left. - \frac{p^2}{(p^2+n)(p^2+n+1)^2(k^2+n)} + \frac{n+1}{(p^2+n+2)(k^2+n)(k^2+n+1)^2}  \right. \nonumber \\
			& \left. - \frac{k^2}{(p^2+n+1)(k^2+n)^2(k^2+n+1)} + \frac{n+1}{(p^2+n+1)^2(p^2+n+2)(k^2+n)} \right] \nonumber \\
			= & -32\pi i N q^4 \sum_{n=0}^{\infty} \int \frac{d^4p d^4k}{(2\pi)^8} \frac{p\cdot k}{(p-k)^2} \nonumber \\
			& \times \left[\frac{n}{(p^2+n)(k^2+n)} - \frac{n+1}{(p^2+n+1)(k^2+n+1)} - \frac{n}{(p^2+n+1)(k^2+n)} \right. \nonumber \\
			& \left. + \frac{n+1}{(p^2+n+2)(k^2+n+1)} + \frac{n}{(p^2+n+1)(k^2+n)^2} - \frac{n+1}{(p^2+n+2)(k^2+n+1)^2} \right] \nonumber \\
		= & \ 0 .
\end{align}

\vspace{\baselineskip}

\begin{figure}[H]
\begin{center}
\includegraphics[width=5cm]{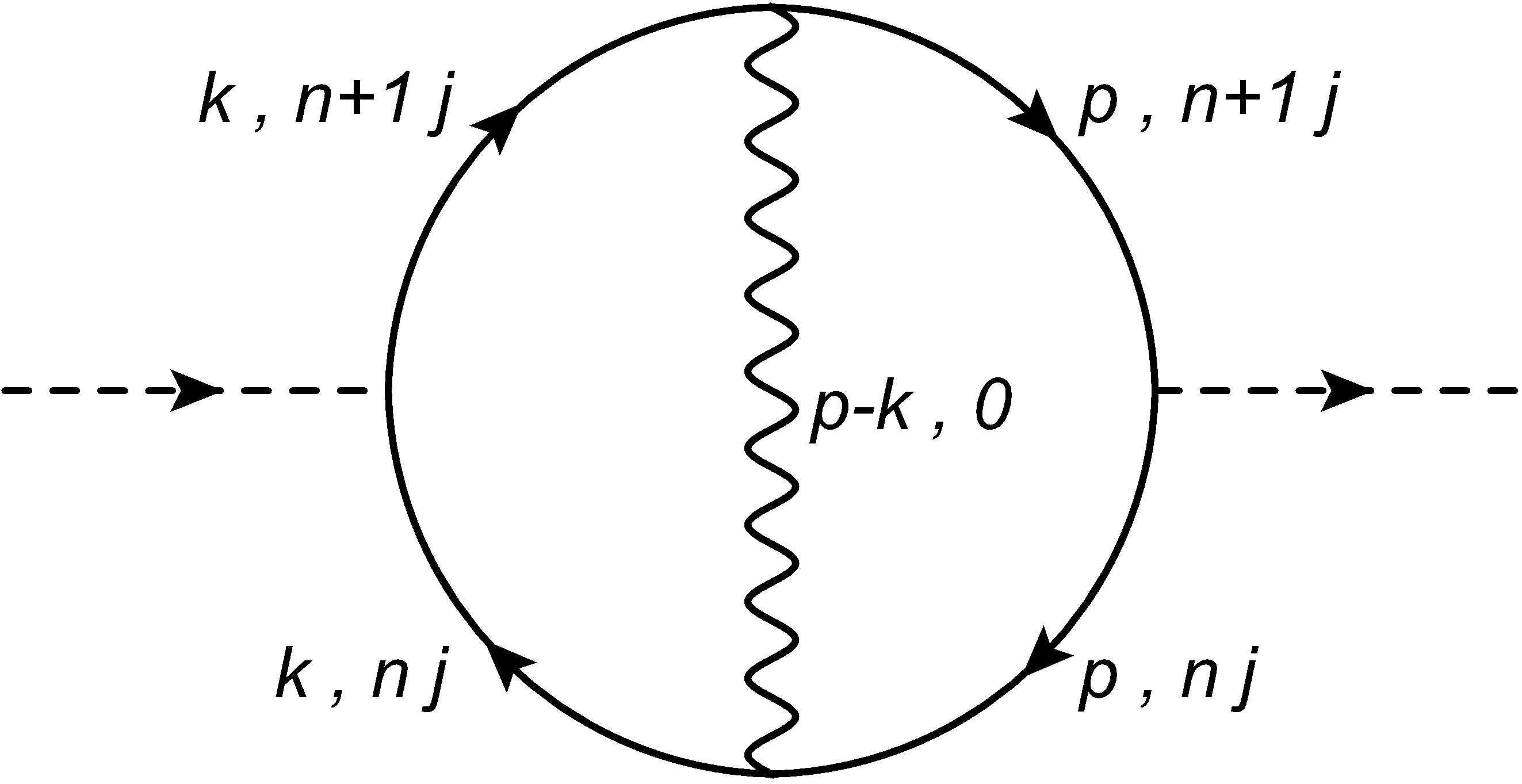} \ \ \ 
\mbox{\raisebox{-3mm}{\includegraphics[width=5cm]{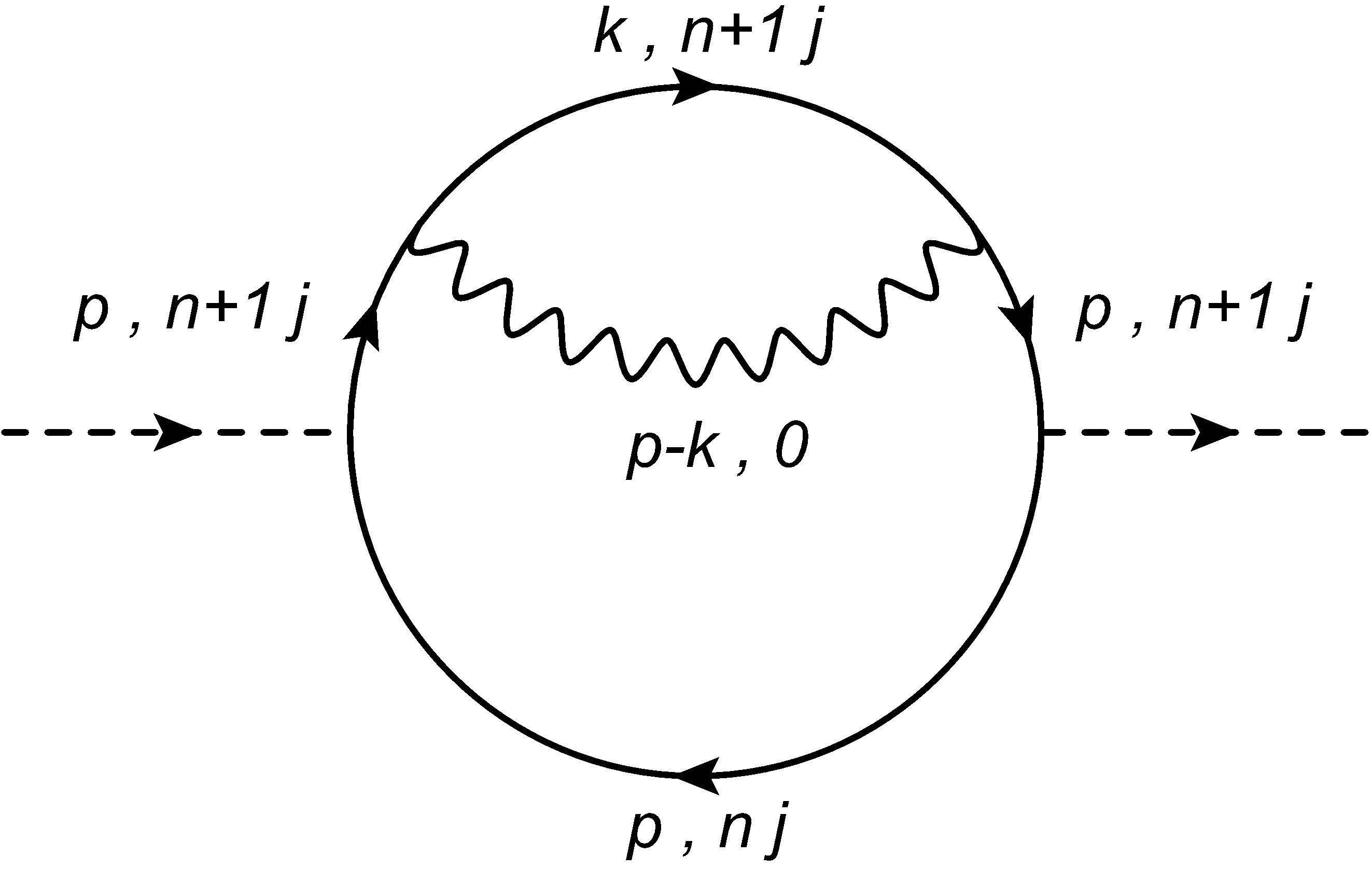}}} \ \ \ 
\mbox{\raisebox{-3mm}{\includegraphics[width=5cm]{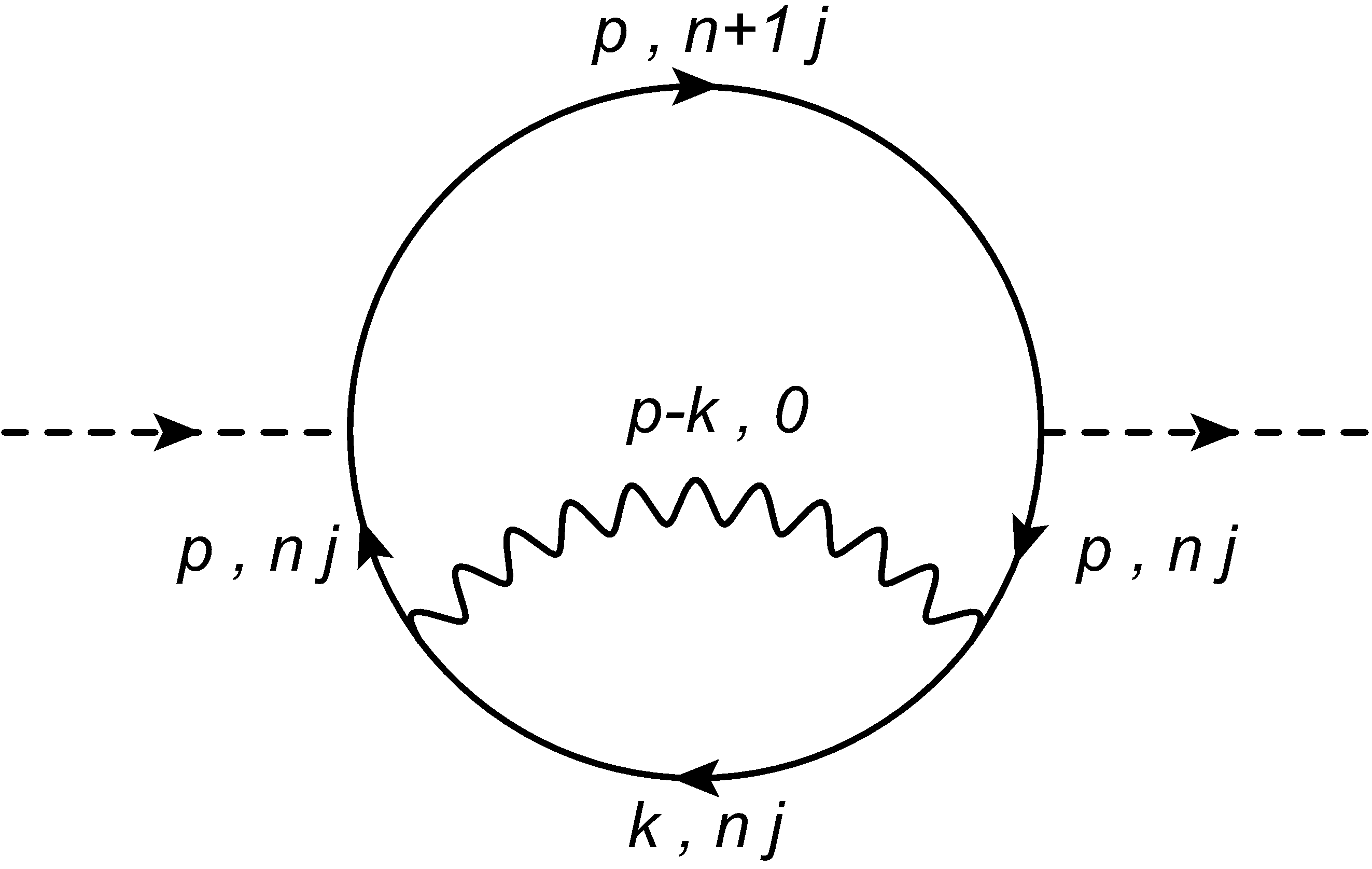}}}
\caption{Two-loop contributions including $A_{\mu,0}$ as an internal line.}
\label{fig.3}
\end{center}
\end{figure}

Next, we consider contributions from the two-loop diagrams including the vector zero mode $A_{\mu,0}$ as an internal line (see fig. \ref{fig.3}). We consider only when the gauge fixing parameter $\xi=1$ for simplicity. The sum of these contributions is also seen to be zero in the same way as above.
\begin{align}
		I^{\mathrm{2 \text{-} loop}}_{A_0} = & -32 \pi i N q^4 \sum_{n=0}^\infty \int \frac{d^4p d^4k}{(2\pi)^8} \frac{1}{(p-k)^2} \nonumber \\
			& \times \left[\frac{(2n+1)p\cdot k + 2p^2k^2 + 2n(n+1)}{(p^2+n)(p^2+n+1)(k^2+n)(k^2+n+1)} \right. \nonumber \\
			& \left. - \frac{(p^2-n-1)p\cdot k +4(n+1)p^2}{(p^2+n)(p^2+n+1)^2(k^2+n+1)} - \frac{(p^2-n)p\cdot k + 4np^2}{(p^2+n)^2(p^2+n+1)(k^2+n)} \right] \nonumber \\
		= & -32\pi i N q^4 \sum_{n=0}^\infty \int \frac{d^4p d^4k}{(2\pi)^8} \frac{1}{(p-k)^2} \nonumber \\
			& \times \left[\left\{ \frac{2n}{(p^2+n)(k^2+n)} - \frac{2(n+1)}{(p^2+n+1)(k^2+n+1)} \right\} p\cdot k - \frac{2n}{(p^2+n)(k^2+n)} \right. \nonumber \\
			& \left. + \frac{2(n+1)}{(p^2+n+1)(k^2+n+1)} + \frac{4n^2}{(p^2+n)^2(k^2+n)} - \frac{4(n+1)^2}{(p^2+n+1)^2(k^2+n+1)} \right] \nonumber \\
		= & \ 0 .
\end{align}

\vspace{\baselineskip}

\begin{figure}[H]
\begin{center}
\includegraphics[width=5cm]{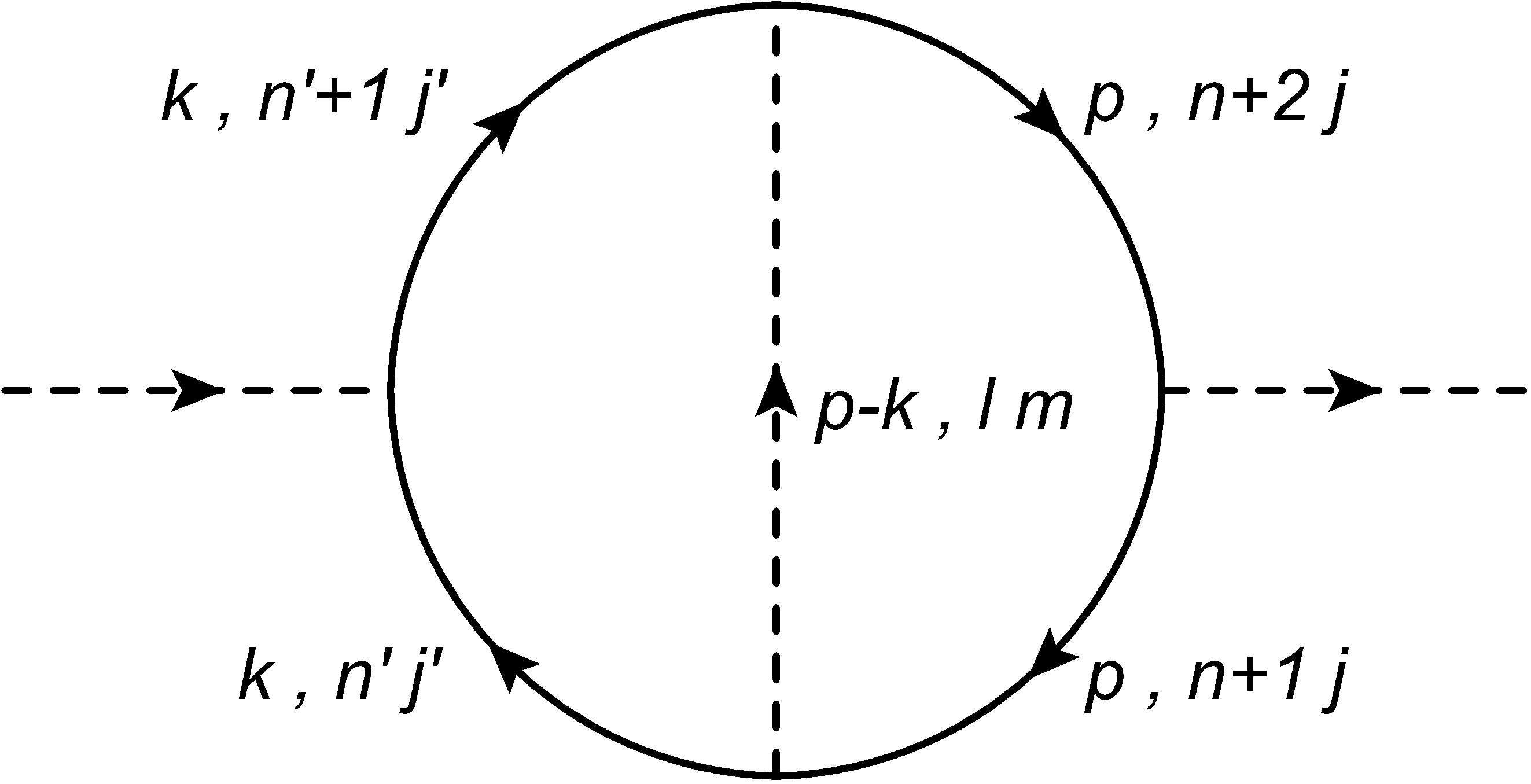} \ \ \ 
\mbox{\raisebox{-3mm}{\includegraphics[width=5cm]{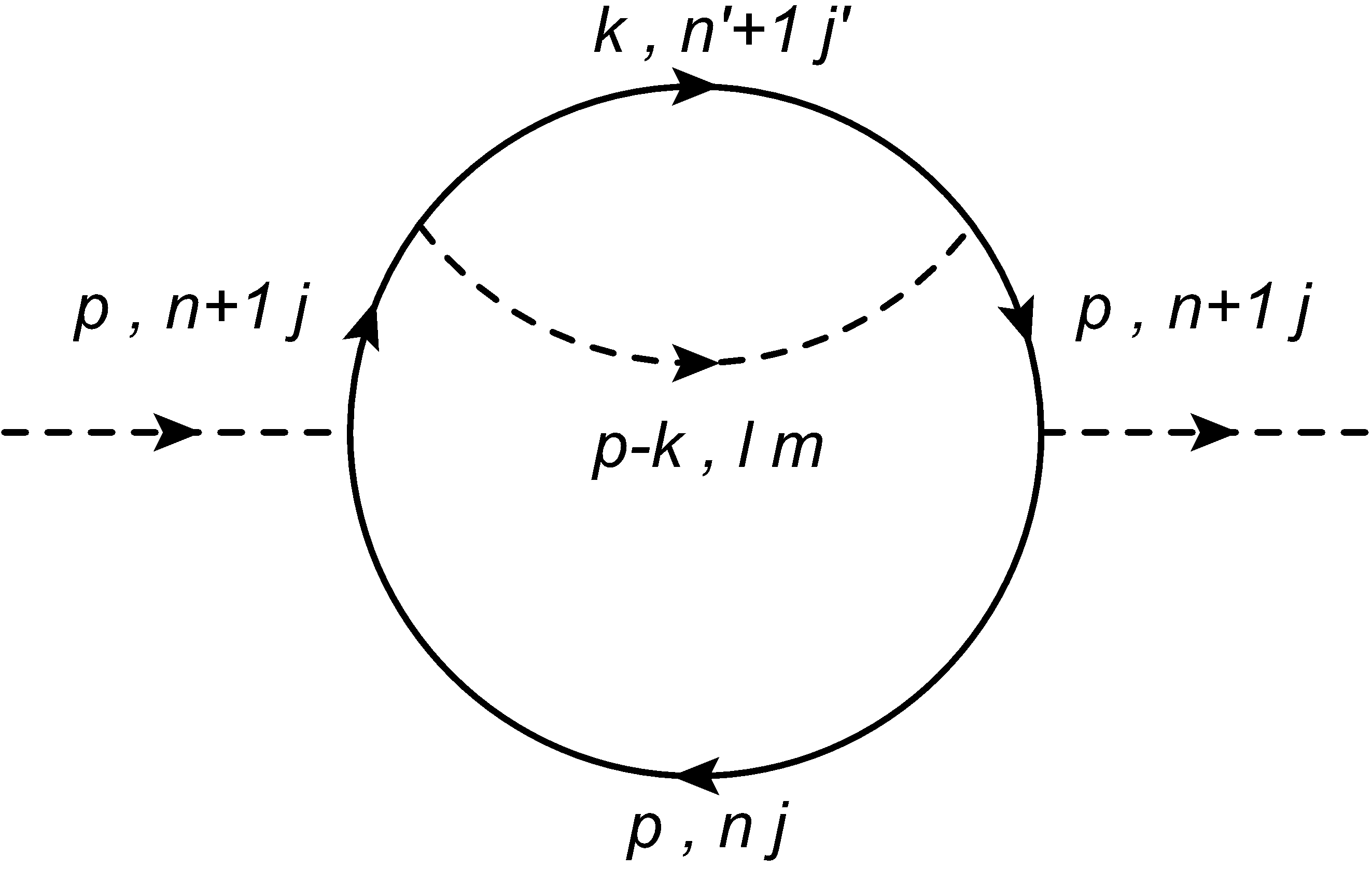}}} \ \ \ 
\mbox{\raisebox{-3mm}{\includegraphics[width=5cm]{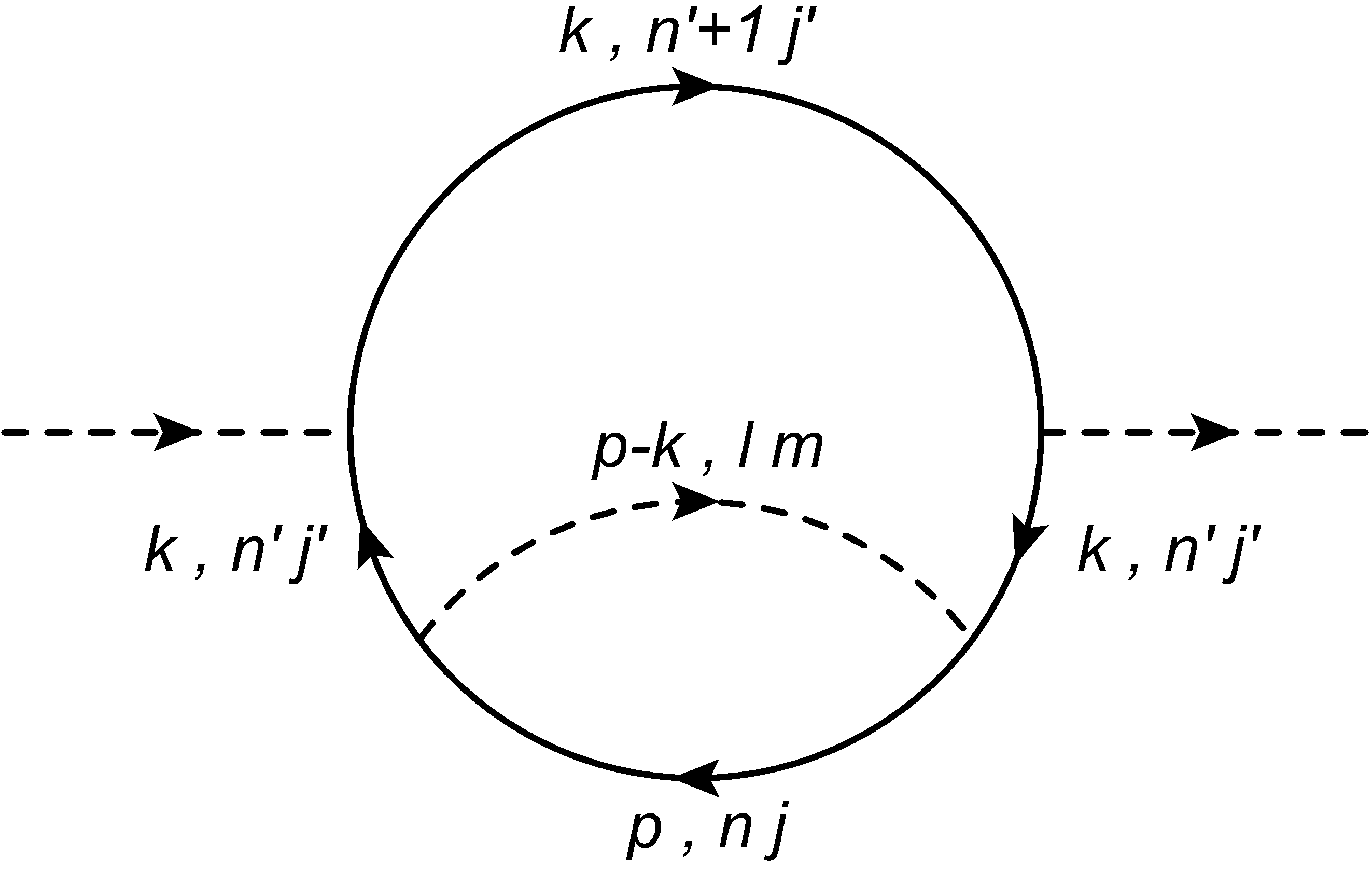}}}
\caption{Two-loop contributions including $\eta_{lm}$ as an internal line.}
\label{fig.4}
\end{center}
\end{figure}

Fig. \ref{fig.4} shows two-loop contributions including massive physical scalars $\eta_{lm}$. We have to note  there are three more types of diagrams where the direction of the scalar internal line is flipped in addition to above three; that is, we consider six types of diagrams in total. The sum over the mode of fermions is
\begin{align}
	I^{\mathrm{2 \text{-} loop}}_{\eta_{lm}} = & - 16\pi i N q^4 \sum_{n,n'=-1}^{\infty} \int \frac{d^4p d^4k}{(2\pi)^8} \frac{1}{(p-k)^2 + \frac{|M_{lm}|^2}{4\pi N}} \nonumber \\
			& \times \left[\frac{\alpha_{nn'} + \beta_{nn'}}{(p^2+n)(k^2+n')} - \frac{\alpha_{nn'+1} + \beta_{nn'}}{(p^2+n)(k^2+n'+1)} - \frac{\alpha_{nn'} + \beta_{n+1n'}}{(p^2+n+1)(k^2+n')} \right. \nonumber \\
			& \left. + \frac{\alpha_{nn'+1} + \beta_{n+1n'}}{(p^2+n+1)(k^2+n'+1)} + \frac{\gamma_{nn'}}{(p^2+n)^2(k^2+n')} - \frac{\gamma_{n+1n'+1}}{(p^2+n+1)^2(k^2+n'+1)} \right. \nonumber \\
			& \left. + \frac{\delta_{nn'}}{(p^2+n)(k^2+n')^2} - \frac{\delta_{n+1n'+1}}{(p^2+n+1)(k^2+n'+1)^2}\right] \nonumber \\
		= & \ 0 ,
\end{align}
where
\begin{align}
	\alpha_{nn'} = & \ - \frac{1}{\sqrt{4\pi N}} \left\{ M_{lm} \left(\sqrt{n}C^{lm}_{n-1j,n'j'} \overline{C}^{lm}_{nj,n'j'} + \sqrt{n+1} C^{lm}_{nj,n'-1j'} \overline{C}^{lm}_{n+1j,n'-1j'} \right) p\cdot k \right. \nonumber \\
			& \ \left. - 2\sqrt{n'} \mathrm{Re} \left(M_{lm} C^{lm}_{nj,n'j'} \overline{C}^{lm}_{nj,n'-1j'} \right) p^2 \right\} ,\\
	\beta_{nn'} = & \ \frac{1}{\sqrt{4\pi N}} \left\{ M_{lm} \left(\sqrt{n'}C^{lm}_{nj,n'j'} \overline{C}^{lm}_{nj,n'-1j'} + \sqrt{n'+1} C^{lm}_{n-1j,n'+1j'} \overline{C}^{lm}_{n-1j,n'j'} \right) p\cdot k \right. \nonumber \\
			& \ \left. - 2\sqrt{n} \mathrm{Re} \left(M_{lm} C^{lm}_{n-1j,n'j'} \overline{C}^{lm}_{nj,n'j'} \right) k^2 \right\}  ,\\
	\gamma_{nn'} = & \ n \left(\left|C^{lm}_{nj,n'-1j'}\right|^2 + \left|C^{lm}_{n-1j,n'j'}\right|^2 \right) p\cdot k - \frac{2\sqrt{nn'}}{|M_{lm}|^2} \mathrm{Re} \left(M^2_{lm} C^{lm}_{n-1j,n'j'} \overline{C}^{lm}_{nj,n'-1j'} \right)p^2 ,\\
	\delta_{nn'} = & \ n' \left(\left|C^{lm}_{nj,n'-1j'}\right|^2 + \left|C^{lm}_{n-1j,n'j'}\right|^2 \right) p\cdot k - \frac{2\sqrt{nn'}}{|M_{lm}|^2} \mathrm{Re} \left(M^2_{lm} C^{lm}_{n-1j,n'j'} \overline{C}^{lm}_{nj,n'-1j'} \right)k^2 .
\end{align}
In this calculation, we use properties of the overlap integral $C^{lm}_{nj,n'j'}$ such as eqs. (\ref{recurrence-1}) and (\ref{recurrence-2}). Note that we define $C^{lm}_{nj,n'j'}=0$ when $n$ or $n'$ is a negative integer.

It is also shown that the sum of two-loop contributions including the NG scalars $\sigma_{lm}$ as an internal line is zero in the almost same way as above. 

\vspace{\baselineskip}

\begin{figure}[H]
\begin{center}
\includegraphics[width=5cm]{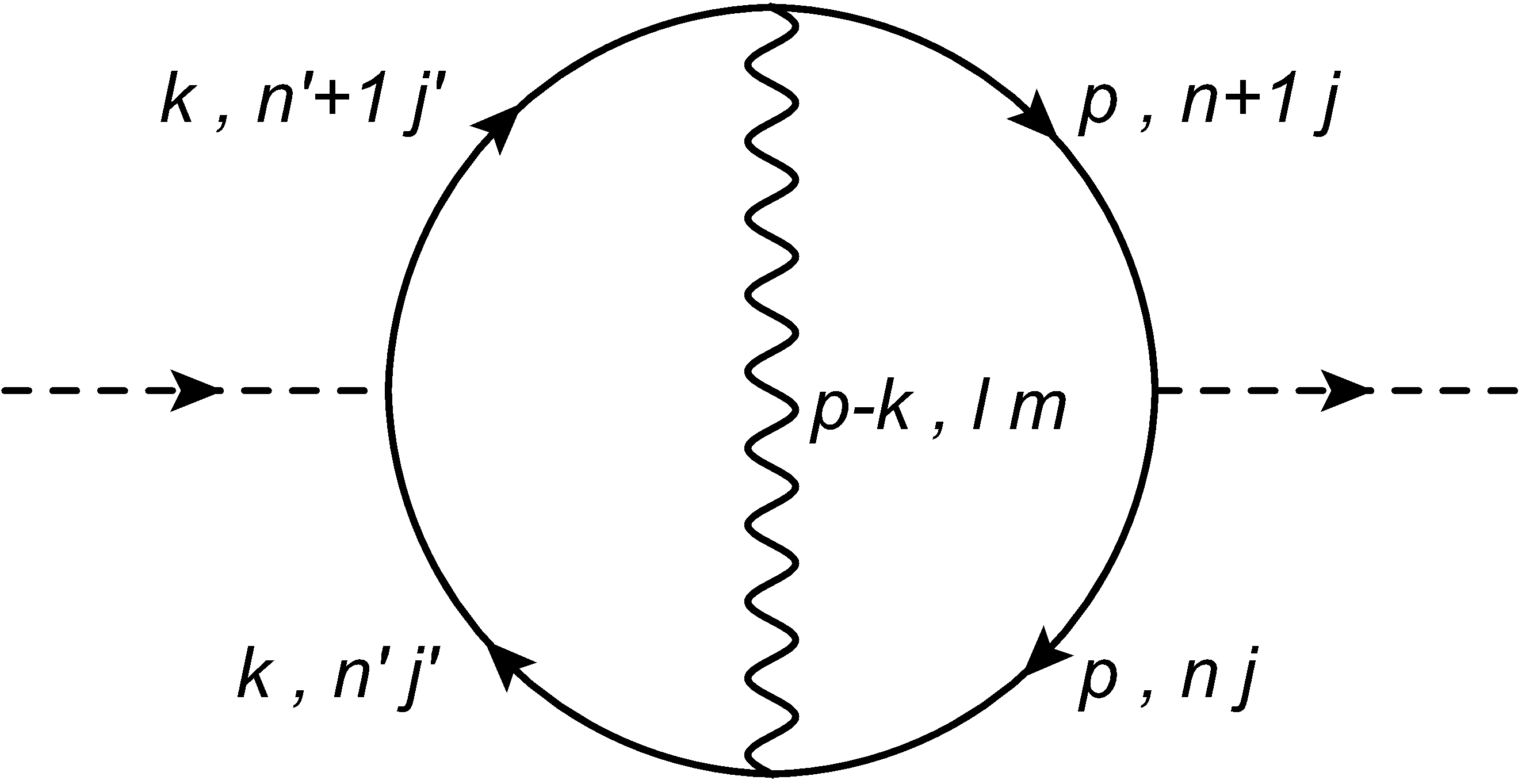} \ \ \ 
\mbox{\raisebox{-3mm}{\includegraphics[width=5cm]{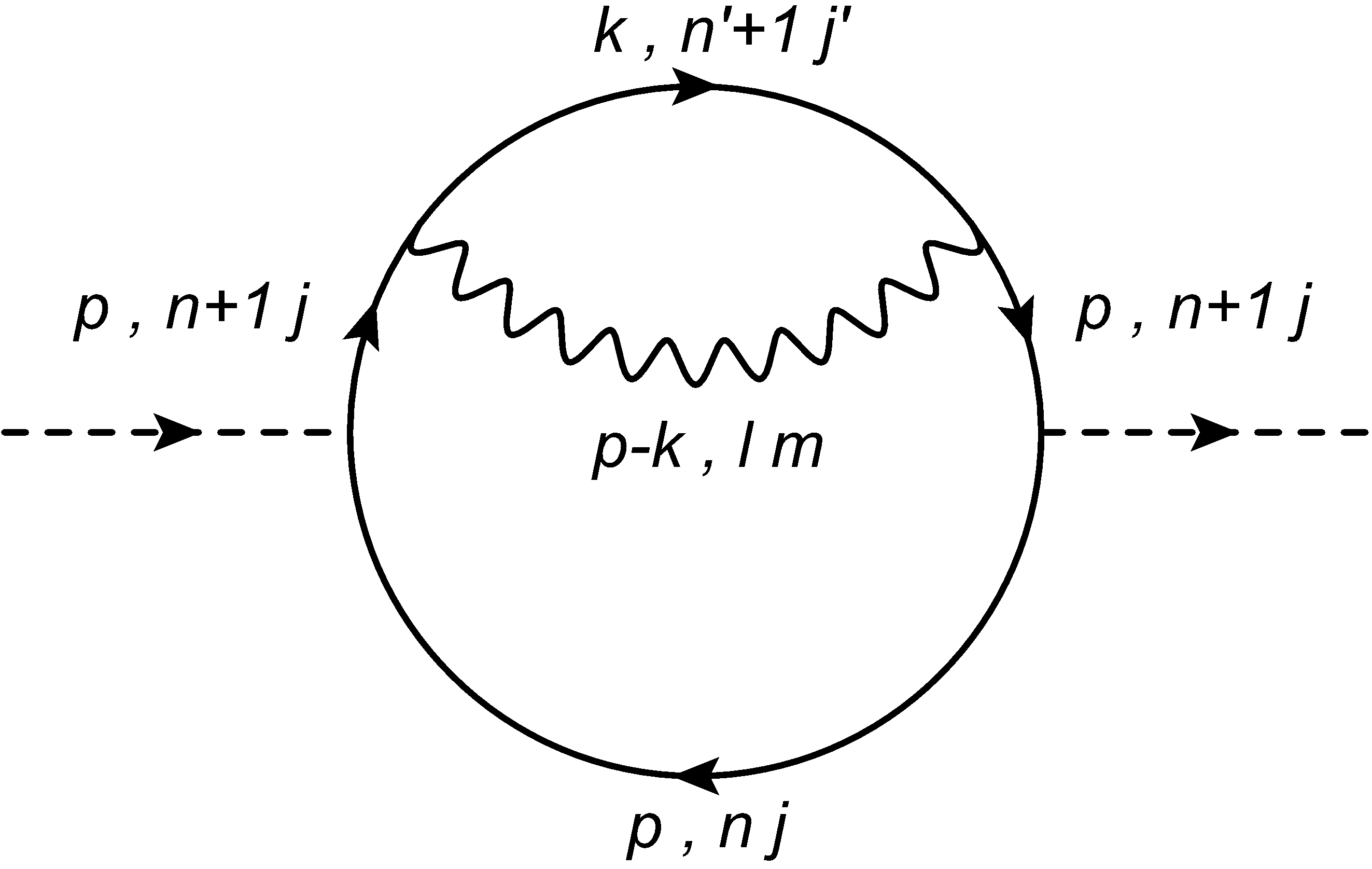}}} \ \ \ 
\mbox{\raisebox{-3mm}{\includegraphics[width=5cm]{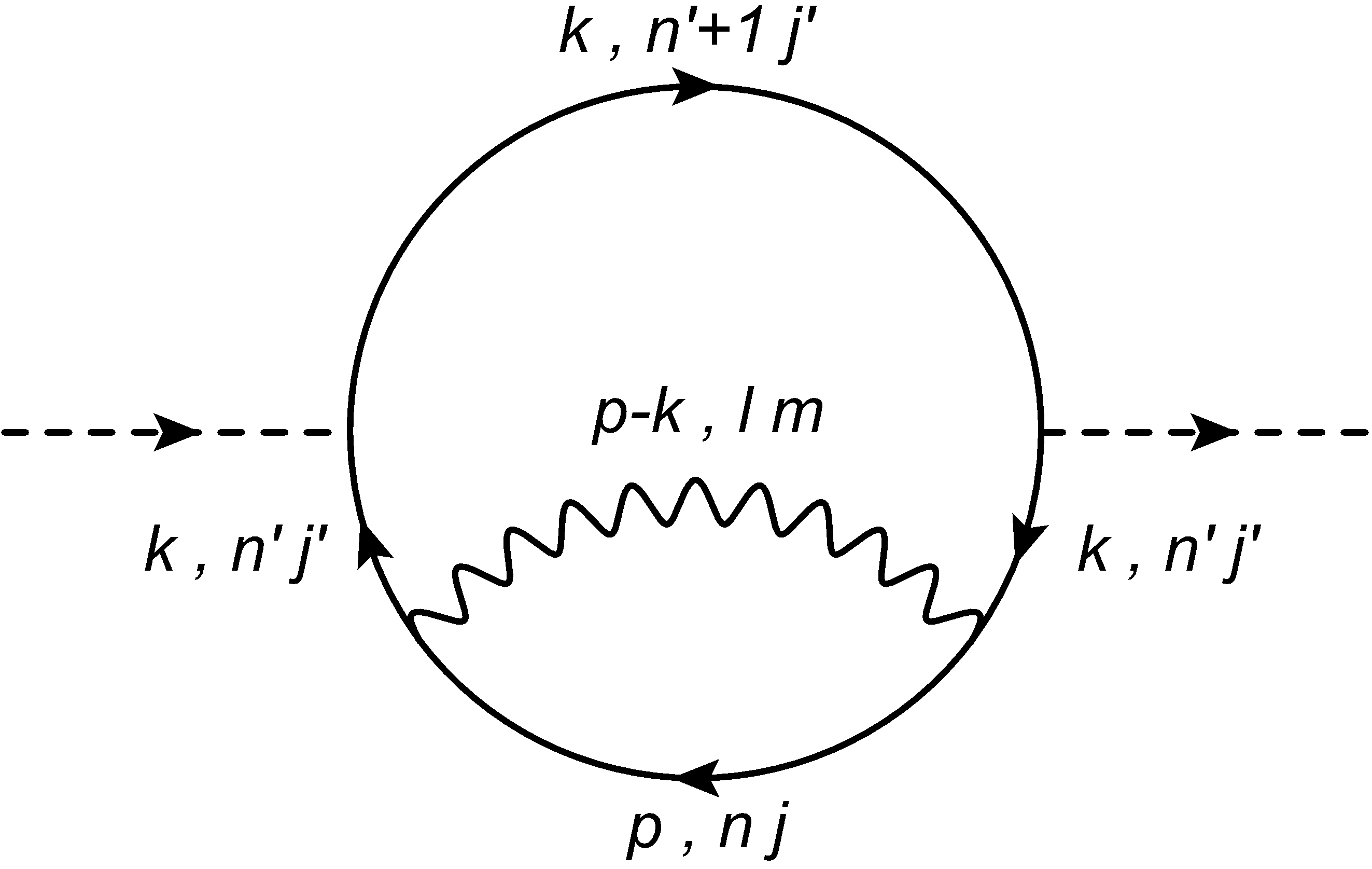}}}
\caption{Two-loop contributions including $A_{\mu,lm}$ as an internal line.}
\label{fig.5}
\end{center}
\end{figure}

Lastly, we consider contributions including massive vectors $A_{\mu,lm}$ (see fig. \ref{fig.5}). We note there are also two choises of the direction of vector internal lines since $A_{\mu,lm}$ is complex. When the gauge fixing parameter $\xi=1$, the sum of them is
\begin{align}
	I^{\mathrm{2 \text{-} loop}}_{A_{lm}} = & -32\pi i N q^4 \sum_{n,n'=-1}^{\infty} \int \frac{d^4p d^4k}{(2\pi)^8} \frac{1}{(p-k)^2 + \frac{\left| M_{lm} \right|^2}{4\pi N}} \nonumber \\
			& \times \left[\frac{\alpha'_{nn'} + \beta'_{nn'} - \epsilon'_{nn'}}{(p^2+n)(k^2+n')} - \frac{\alpha'_{nn'+1} + \delta'_{nn'+1}}{(p^2+n)(k^2+n'+1)} - \frac{\beta'_{n+1n'} + \gamma'_{n+1n'}}{(p^2+n+1)(k^2+n')} \right. \nonumber \\
			& \left. + \frac{\gamma'_{n+1n'+1} + \delta'_{n+1n'+1} + \epsilon'_{n+1n'+1}}{(p^2+n+1)(k^2+n'+1)} + \frac{\zeta'_{nn'}}{(p^2+n)^2(k^2+n')} - \frac{\zeta'_{n+1n'+1}}{(p^2+n+1)^2(k^2+n'+1)} \right. \nonumber \\
			& \left. + \frac{\eta'_{nn'}}{(p^2+n)(k^2+n')^2} - \frac{\eta'_{n+1n'+1}}{(p^2+n+1)(k^2+n'+1)^2}\right] \nonumber \\
		= & \ 0 ,
\end{align}
where
\begin{align}
	\alpha'_{nn'} =& \ - \frac{M_{lm}}{\sqrt{4\pi N}} \left\{ \left(\sqrt{n} C^{lm}_{n-1j,n'-1j'} \overline{C}^{lm}_{nj,n'-1j'} + \sqrt{n+1} C^{lm}_{nj,n'j'} \overline{C}^{lm}_{n+1j,n'j'} \right) p\cdot k \right. \nonumber \\
			& \ \left. + 2\sqrt{nn'} \left(\sqrt{n} C^{lm}_{nj,n'j'} \overline{C}^{lm}_{nj,n'-1j'} + \sqrt{n+1} C^{lm}_{n-1j,n'-1j'} \overline{C}^{lm}_{n+1j,n'j'} \right) \right\} ,\\
	\beta'_{nn'} =& \ \frac{M_{lm}}{\sqrt{4\pi N}} \left\{ \left(\sqrt{n'} C^{lm}_{n-1j,n'j'} \overline{C}^{lm}_{n-1j,n'-1j'} + \sqrt{n'+1} C^{lm}_{nj,n'+1j'} \overline{C}^{lm}_{nj,n'j'} \right) p\cdot k \right. \nonumber \\
			& \ \left. + 2\sqrt{nn'} \left(\sqrt{n'} C^{lm}_{n-1j,n'j'} \overline{C}^{lm}_{nj,n'j'} + \sqrt{n'+1} C^{lm}_{nj,n'+1j'} \overline{C}^{lm}_{n-1j,n'-1j'} \right) \right\} ,\\
	\gamma'_{n+1n'} =& \ - \frac{M_{lm}}{\sqrt{4\pi N}} \left\{ \left(\sqrt{n} C^{lm}_{n-1j,n'-1j'} \overline{C}^{lm}_{nj,n'-1j'} + \sqrt{n+1} C^{lm}_{nj,n'j'} \overline{C}^{lm}_{n+1j,n'j'} \right) p\cdot k \right. \nonumber \\
			& \ \left. + 2\sqrt{(n+1)n'} \left(\sqrt{n} C^{lm}_{n-1j,n'-1j'} \overline{C}^{lm}_{n+1j,n'j'} + \sqrt{n+1} C^{lm}_{nj,n'j'} \overline{C}^{lm}_{nj,n'-1j'} \right) \right\} ,\\
	\delta'_{nn'+1} =& \ \frac{M_{lm}}{\sqrt{4\pi N}} \left\{ \left(\sqrt{n'} C^{lm}_{n-1j,n'j'} \overline{C}^{lm}_{n-1j,n'-1j'} + \sqrt{n'+1} C^{lm}_{nj,n'+1j'} \overline{C}^{lm}_{nj,n'j'} \right) p\cdot k \right. \nonumber \\
			& \ \left. + 2\sqrt{n(n'+1)} \left(\sqrt{n'} C^{lm}_{nj,n'+1j'} \overline{C}^{lm}_{n-1j,n'-1j'} + \sqrt{n'+1} C^{lm}_{n-1j,n'j'} \overline{C}^{lm}_{nj,n'j'} \right) \right\} ,\\
	\epsilon'_{nn'} =& \ 4\sqrt{nn'} \mathrm{Re}\left(C^{lm}_{nj,n'j'} \overline{C}^{lm}_{n-1j,n'-1j'} \right) ,\\
	\zeta'_{nn'} =& \  n \left(\left|C^{lm}_{n-1j,n'-1j'}\right|^2 + \left|C^{lm}_{nj,n'j'}\right|^2 \right) p\cdot k + 4n\sqrt{nn'} \mathrm{Re} \left(C^{lm}_{nj,n'j'} \overline{C}^{lm}_{n-1j,n'-1j'} \right) ,\\
	\eta'_{nn'} =& \  n' \left(\left|C^{lm}_{n-1j,n'-1j'}\right|^2 + \left|C^{lm}_{nj,n'j'}\right|^2 \right) p\cdot k + 4n' \sqrt{nn'} \mathrm{Re} \left(C^{lm}_{nj,n'j'} \overline{C}^{lm}_{n-1j,n'-1j'} \right) .
\end{align}
Therefore, quantum corrections to the mass of the WL scalar vanish up to the two-loop level.

\subsection{Effective Potential}

The WL phase with an appropriate factor $f \overline{\theta}/\sqrt{2}$ is a VEV of the WL scalar $\varphi_0$. Thus, its mass squared $m^2_{\varphi_0}$ is obtained as the second derivative of the effective potential $V_{\mathrm{eff}}$ in the 4D effective theory respect to the VEV,
\begin{align}
	m^2_{\varphi_0} = \frac{2}{f^2} \frac{\partial^2 V_{\mathrm{eff}}}{\partial \theta \partial \overline{\theta}} .
\end{align}
In our setup, propagators or mass spectra in the 4D effective theory are independent of the WL phase $\theta$. Furthermore, bubble graphs which contribute to $V_{\mathrm{eff}}$ always include conjugate pare vertices proportional to $e^{\theta M_{lm} - \overline{\theta} \overline{M}_{lm}}$ or $e^{-\theta M_{lm} + \overline{\theta} \overline{M}_{lm}}$. Thus, $\theta$-dependence of vertices is cancelled in each bubble graph and $V_{\mathrm{eff}}$ is independent of $\theta$. Therefore, the WL scalar is massless in the full order.

Because of $\theta$-independence of $V_{\mathrm{eff}}$, a mechanism which dynamically determines the value of the WL phase, i.e. the Hosontani mechanism \cite{Hosotani:1983xw, Hosotani:1988bm}, does not work in our setup.

\section{Summary}
\label{sec.5}

In this paper, we have studied properties of the WL scalar in a 6D U(1) gauge theory compactified on a torus with magnetic flux. In the previous researches, it has been shown that the quantum corrections to the masses of the WL scalars vanish at the one-loop level in a few gauge theories on a magnetized torus. The physical reason for this masslessness has been discussed to be the shift symmetry of the WL scalars or that they are NG bosons of the translations on the torus. 

First, we derived the 4D effective Lagrangian including the WL phase from the dimensional reduction of U(1) gauge theory on $M^4 \times T^2$ with magnetic flux. The main target of this paper was a scalar zero mode originated from the extradimensional components of the gauge field.

The translational symmetry on the torus is breaking spontaneously due to magnetic flux. To confirm the WL scalar is an NG boson of this translations, we computed commutation relations between the WL scalar and the momentum operators. In the 6D theory, extradimensional components of the gauge field are NG bosons and their property as NG bosons is taken over by the WL scalar after dimensional reduction.

Using the 4D effective Lagrangian, we have calculated quantum corrections to the mass of the WL scalar up to the two-loop level. The corrections cancel non-trivially by summing up contributions from all the fermion KK modes. To confirm the masslessness of the WL scalar, we have also discussed the effective potential. Since it is independent of the WL phase, the WL scalar is indeed massless in the full order. In addition, the WL phase is not determined dynamically like some gauge-Higgs unification (GHU) models.

In this paper, we have focused only on a 6D U(1) gauge theory compactified on a magnetized torus. It would be interesting to confirm the cancellation of higher-loop corrections in other gauge theories or to generalize this result to flux compactifications on other manifolds.

Since massless scalars are phenomenologically undesirable, we are also interested in how to make the WL scalars massive, that is, pseudo-NG bosons without spoiling desirable results of flux compactifications such as the chiral structure and the generation structure of fermions. If it is possible, the application to GHU models might be interesting.

\section*{Acknowledgments}

We would like to thank Hiroyuki Abe for useful comments and discussions.

\appendix

\section{Gamma Matrices}
\label{app.A}
The gamma matrices we used in this paper satisfy the 6D Clifford algebra
\begin{align}
	\{\Gamma^M , \Gamma^N\} = -2g^{MN} .
\end{align}
Our choice for 6D gamma matrices are
\begin{align}
	\Gamma^\mu = \left(\begin{array}{cc} \gamma^\mu & 0 \\ 0 & \gamma^\mu \end{array} \right) ,\ \ \ \ \ \ \ \ 
	\Gamma^5 = \left(\begin{array}{cc} 0 & i\gamma^5 \\ i\gamma^5 & 0 \end{array} \right) ,\ \ \ \ \ \ \ \ 
	\Gamma^6 = \left(\begin{array}{cc} 0 & -\gamma^5 \\ \gamma^5 & 0 \end{array} \right) , 
\end{align}
where the 4D gamma matrices are given by
\begin{align}
	\gamma^\mu = \left(\begin{array}{cc} 0 & \sigma^\mu \\ \overline{\sigma}^\mu & 0 \end{array} \right) ,\ \ \ \ \ \ \ \ 
	\gamma^5 = i\gamma^0\gamma^1\gamma^2\gamma^3 = \left(\begin{array}{cc} -1 & 0 \\ 0 & 1 \end{array} \right) .
\end{align}

In this notation, the 6D chirality operator is given by
\begin{align}
	\Gamma^7 = \Gamma^0\Gamma^1\Gamma^2\Gamma^3\Gamma^5\Gamma^6 = \left(\begin{array}{cc} \gamma^5 & 0 \\ 0 & -\gamma^5 \end{array} \right) .
\end{align}
For these gamma matrices, we can decompose a left-handed 6D Weyl fermion $\Psi$ into two-component Weyl fermions $\psi$ and $\chi$ as follows.
\begin{align}
	\Psi = \left(\begin{array}{c} \psi_L \\ \psi_R \end{array} \right) : \ \ \ \ \ \ \ \ 
	&\gamma_5\psi_L = -\psi_L ,\ \ \ \ \ \ \ \ 
	\gamma_5\psi_R = \psi_R , \nonumber \\ 
	&\psi_L = \left(\begin{array}{c} \psi \\ 0 \end{array} \right) ,\ \ \ \ \ \ \ \ 
	\psi_R = \left(\begin{array}{c} 0 \\ \chi^\dagger \end{array} \right) .
\end{align}


\bibliographystyle{prsty}

\end{document}